\documentclass[10pt,journal,compsoc]{IEEEtran}

\usepackage{color}
\usepackage[nocompress]{cite}
\usepackage[pdftex]{graphicx}
\graphicspath{{../pdf/}{../jpeg/}}
\DeclareGraphicsExtensions{.pdf,.jpeg,.png}
\usepackage[caption=false,font=footnotesize,labelfont=sf,textfont=sf]{subfig}

\usepackage{amsmath}
\usepackage{amsthm}
\usepackage{amssymb}
\usepackage{bm}
\usepackage{latexsym}
\usepackage{float}
\usepackage{booktabs}
\usepackage{algpseudocode}
\usepackage{algorithmicx,algorithm}
\usepackage{array}
\usepackage{mdwmath}
\usepackage{mdwtab}
\usepackage{eqparbox}
\usepackage{mathtools}
\usepackage{url}
\usepackage{enumerate}
\newcommand{\sectionref}[1]{Section \ref{#1}}
\newcommand{\figref}[1]{Fig. \ref{#1}}
\newcommand{\tabref}[1]{Table \ref{#1}}
\newcommand{\eqaref}[1]{Eq. (\ref{#1})}

\theoremstyle{definition}
\newtheorem{remark}{Remark} 

\theoremstyle{definition}

\ifCLASSOPTIONcompsoc
  \usepackage[nocompress]{cite}
\else
  \usepackage{cite}
\fi
\ifCLASSINFOpdf
\else
\fi
\begin{document}
%
 \title{UAV-enabled Collaborative Beamforming via Multi-Agent Deep Reinforcement Learning}
%
%
%
%

\author{Saichao Liu,
	Geng Sun*,~\IEEEmembership{Member,~IEEE,}
        Jiahui Li*,~\IEEEmembership{Student Member,~IEEE,}
        Shuang Liang,\\
	Qingqing Wu,~\IEEEmembership{Senior Member,~IEEE,}
        Pengfei Wang,~\IEEEmembership{Member,~IEEE,}
	and Dusit Niyato,~\IEEEmembership{Fellow,~IEEE}
	
	\thanks{
	\par This study is supported in part by the National Natural Science Foundation of China (62272194, 62172186), and in part by the Science and Technology Development Plan Project of Jilin Province (20230201087GX). \textit{(Corresponding authors: Geng Sun and Jiahui Li.)}
 
 \par Saichao Liu is with the College of Computer Science and Technology, Jilin University, Changchun 130012, China (e-mail: saichao22@mails.jlu.edu.cn).

	\par Geng Sun is with the College of Computer Science and Technology, Jilin University, Changchun 130012, China, and with Key Laboratory of Symbolic Computation and Knowledge Engineering of Ministry of Education, Jilin University, Changchun 130012, China; he is also affiliated with the College of Computing and Data Science, Nanyang Technological University, Singapore 639798 (e-mail: sungeng@jlu.edu.cn).
	
	\par Jiahui Li is with the College of Computer Science and Technology, Jilin University, Changchun 130012, China (e-mail: lijiahui0803@foxmail.com).
 
     \par Shuang Liang is with the School of Information Science and Technology, Northeast Normal University, Changchun 130117, China (e-mail: liangshuang@nenu.edu.cn).
	
	\par Qingqing Wu is with the Department of Electronic Engineering, Shanghai Jiao Tong University, Shanghai 200240, China (e-mail: qingqingwu@sjtu.edu.cn).

        \par Pengfei Wang is with the School of Computer Science and Technology, Dalian University of Technology, Dalian 116024, China (e-mail: wangpf@dlut.edu.cn).
	
	\par Dusit Niyato is with the College of Computing and Data Science, Nanyang Technological University, Singapore 639798 (email: dniyato@ntu.edu.sg). 
\protect}
	}

%
%

\markboth{Journal of \LaTeX\ Class Files,~Vol.~14, No.~8, August~2015}%
{Shell \MakeLowercase{\textit{et al.}}: Bare Advanced Demo of IEEEtran.cls for IEEE Computer Society Journals}
%



\IEEEtitleabstractindextext{%
\begin{abstract}
In this paper, we investigate an unmanned aerial vehicle (UAV)-assistant air-to-ground communication system, where multiple UAVs form a UAV-enabled virtual antenna array (UVAA) to communicate with remote base stations by utilizing collaborative beamforming. To improve the work efficiency of the UVAA, we formulate a UAV-enabled collaborative beamforming multi-objective optimization problem (UCBMOP) to simultaneously maximize the transmission rate of the UVAA and minimize the energy consumption of all UAVs by optimizing the positions and excitation current weights of all UAVs. This problem is challenging because these two optimization objectives conflict with each other, and they are non-concave to the optimization variables. Moreover, the system is dynamic,  and the cooperation among UAVs is complex, making traditional methods take much time to compute the optimization solution for a single task. In addition, as the task changes, the previously obtained solution will become obsolete and invalid. To handle these issues, we leverage the multi-agent deep reinforcement learning (MADRL) to address the UCBMOP. Specifically, we use the heterogeneous-agent trust region policy optimization (HATRPO) as the basic framework, and then propose an improved HATRPO algorithm, namely HATRPO-UCB, where three techniques are introduced to enhance the performance. Simulation results demonstrate that the proposed algorithm can learn a better strategy compared with other methods. Moreover, extensive experiments also demonstrate the effectiveness of the proposed techniques.
\end{abstract}

}

\maketitle

\IEEEdisplaynontitleabstractindextext

%
\IEEEpeerreviewmaketitle

\ifCLASSOPTIONcompsoc
\IEEEraisesectionheading{\section{Introduction}\label{sec:introduction}}
\else
\section{Introduction}
\label{sec:introduction}
\fi
\par \IEEEPARstart{U}{nmanned} aerial vehicles (UAVs) are aircraft without pilots, which are flying by remote or autonomous control \cite{zeng2019accessing}. With the advantages of high mobility, low cost and line-of-sight (LoS) communications, UAVs can be applied in many wireless communication scenarios. For instance, a UAV can act as an aerial base station (BS) to provide emergency communication service for ground users when the ground BSs are unavailable\cite{zhao2019uav,Chen2021}. Moreover, a UAV can serve as a relay to improve the coverage and connectivity of wireless networks \cite{ahmed2020energy}. In addition, the UAVs can be regarded as the new aerial users which is able to deliver packages~\cite{khosravi2020unmanned}.

\par Despite the application prospect of UAVs being promising, many challenges still exist. Specifically, the limited battery capacity will affect the work efficiency of a UAV \cite{zeng2016wireless} \cite{li2021minimizing}, thus some relevant optimizations (e.g., flight trajectory \cite{zeng2019energy} and mission execution time \cite{zhan2022energy}) are necessary to prolong the working time of UAVs. Besides, a UAV is difficult to communicate with remote BSs due to its limited transmission power \cite{mozaffari2019tutorial}, and it cannot fly close to the BSs because of the finite energy capacity. Therefore, it is significant to improve the transmission ability of a UAV. 

\par Collaborative beamforming (CB) has been justified to efficiently improve the transmission performance of distributed systems with limited resources \cite{liang2020joint}. Specifically, the UAVs can form a virtual antenna array (VAA), where each UAV is regarded as a single antenna element, then the UAV-assisted VAA (UVAA) can produce a high-gain mainlobe toward the remote BS by utilizing CB, such that achieving the long distance air-to-ground (A2G) communication between the UAVs and remote BS. To obtain an optimal beam pattern so that achieving a higher receiving rate of the BS, all UAVs need to fly to more appropriate positions and adjust their excitation current weights \cite{garza2016design}. However, for communicating with different BSs, all UAVs need to adjust their positions constantly, which will consume extra energy. Hence, in addition to improving the transmission rate, the optimization to the motion energy consumption of UVAA is also required so as to prolong the lifespan of UVAA. 

\par Previous studies usually optimize the communication performance or energy efficiency separately by using the traditional methods (e.g., convex optimization, or evolutionary computation). However, it is difficult to achieve optimal performance of these two optimization objectives simultaneously because they conflict with each other. Furthermore, since the cooperation among UAVs is very complex and the number of UAVs is usually large in the UVAA system, traditional methods will take much time to compute the optimization results. Besides, the system is dynamic, causing the currently obtained optimization results will become invalid when the system status changes. Therefore, it is inappropriate to apply traditional methods to simultaneously optimize the two objectives of the UVAA system.

\par Deep reinforcement learning (DRL) is an effective approach to solve complex and dynamic optimization problems without prior knowledge, and it has been proven to be an effective method to deal with complex optimization problems with high-dimensional continuous spaces \cite{fujimoto2018addressing}. Several studies have applied DRL to solve single-UAV communication optimization problems \cite{wang2021trajectory} \cite{xie2021connectivity}. However, DRL methods usually consider only single-agent learning frameworks, which are inappropriate for dealing with the UAV-enabled CB optimization problem. Thus, multi-agent deep reinforcement learning (MADRL) can be regarded as a potential approach to obtain the optimal policies since the learned policy of each agent involves the cooperation with other agents in MADRL. Specifically, all agents use the deep neural network (DNN) for policy learning. Benefiting from the powerful learning ability of DNN, all agents can learn excellent strategies. Meanwhile, compared with traditional methods, all agents can make corresponding decision directly without consuming much computing resource when the environment changes.

\par In this work, we utilize MADRL to deal with the UAV-enabled CB optimization problem. Consider that the cooperation is very complex among UAVs, we use the heterogeneous-agent trust region policy optimization (HATRPO) \cite{kuba2021trust} as the main MADRL framework. Besides, the number of UAVs is usually large in the UVAA, which may cause the curse of dimensionality for MADRL algorithm. Furthermore, the policy gradient estimation problem may exists. Therefore, we propose an improved algorithm for solving our constructed problem well. The main contributions are summarized as follows:

\begin{itemize}
	\item \textit{\textbf{UAV-enabled CB Multi-objective Optimization Formulation:}} We study a UAV-assistant A2G communication system, where a swarm of UAVs form a UVAA to perform CB for communicating with remote BSs. Then, we formulate a UAV-enabled CB multi-objective optimization problem (UCBMOP), aiming at simultaneously maximizing the transmission rate of UVAA and minimizing the energy consumption of the UAV elements by optimizing the positions and the excitation current weights of all UAVs. The problem is difficult to address because the two optimization objectives conflict with each other, and they are non-concave with respect to the optimization variables, such as the coordinates and the excitation current weights of UAVs.

	\item \textit{\textbf{New MADRL Approach:}} We model a multi-agent Markov decision process, called Markov game, for characterizing the formulated UCBMOP. Then, we propose an improved HATRPO algorithm, namely HATRPO-UCB, to solve the UCBMOP. Specifically, three techniques are proposed to enhance the performance of conventional HATRPO, which are  \emph{observation enhancement}, \emph{agent-specific global state} and \emph{Beta distribution for policy}, respectively. The first technique is to combine the algorithm with the problem well, so as to learn the better strategy for UCBMOP. The other two techniques are used to enhance the learning performance of critic and actor network, respectively. 
	
	\item \textit{\textbf{Simulation and Performance Analysis:}} Simulation results show that the proposed HATRPO-UCB learns the best strategy compared with two classic antenna array solutions, three baseline MADRL algorithms and the conventional HATRPO. Besides, ablation experiments also demonstrate the effectiveness of our proposed techniques.
	
\end{itemize}

\par The rest of the paper is organized as follows. \sectionref{sec:related_work} discusses the related works. The system model and problem formulation are presented in \sectionref{sec:system_model_and_problem_formulation}. \sectionref{sec:HATRPO_UCB} proposes HATRPO-UCB. Simulation results and analysis are provided in \sectionref{sec:simulation_results_and_analysis}. Finally, \sectionref{sec:conclusion} concludes the paper.

\section{Related Work}
\label{sec:related_work}
\par Several previous works have studied the energy optimization in UAV-based communication scenarios. For example, Zeng \textit{et al.} \cite{zeng2019energy} considered a scenario where a UAV needs to communicate with multiple ground nodes, then proposed an efficient algorithm to achieve the energy minimization by jointly optimizing the hovering locations and durations, as well as the flying trajectory connecting these hovering locations. Cai \textit{et al.} \cite{cai2020joint} studied the energy-efficient secure downlink UAV communication problem, with the purpose of maximizing the system energy efficiency. The authors divided the optimization problem into two subproblems, and proposed an alternating optimization-based algorithm to jointly optimize the resource allocation strategy, the trajectory of information UAV, and the jamming policy of jammer UAV. Wang \textit{et al.} \cite{wang2023stackelberg} investigated a multiple UAVs-assisted mobile edge computing (MEC) system, where a natural disaster has damaged the edge server. The authors proposed two game theory-based algorithms, where the first algorithm is used to minimize the total energy consumption of multiple UAVs, and the second one is designed to achieve the utility trade-off between the UAV-MEC server and mobile users. 

\par Except for adopting traditional methods,  some works also use DRL to optimize the energy consumption of UAV. For optimizing the energy consumption of cellular-connected UAV, Zhan \textit{et al.} \cite{zhan2022energy} developed a DRL algorithm based on dueling deep Q network (DQN) with multi-step double Q-learning to jointly optimize the mission completion time, trajectory of UAV and associations to BSs. Abedin \textit{et al.} \cite{abedin2020data} propose an agile deep reinforcement
learning algorithm to achieve an energy-efficient trajectory optimization for
multiple UAVs so as to improve the data freshness and
connectivity to the Internet of Things (IoT) devices. Liu \textit{et al.} \cite{liu2019distributed} proposed a decentralized DRL-based framework to control multiple UAVs for achieving the long-term communication coverage in an area. The goal is to simultaneously maximize the temporal average coverage score achieved by all UAVs in a task, maximize the geographical fairness of all considered point-of-interests (PoIs), and minimize the total energy consumption. 

\par UAV communications enabled by CB are also investigated in previous works. For example, Garza \textit{et al.} \cite{garza2016design} proposed the differential evolution method for a UAV-based 3D antenna array, aiming at achieving a maximum performance with respect to directivity and sidelobe level (SLL). Mozaffari \textit{et al.} \cite{mozaffari2018communications} considered a UAV-based linear antenna array (LAA) communication system and proposed two optimization algorithms, where the first algorithm is used to minimize the transmission time for ground users, and the second one is designed to optimize the control time of UAV-based LAA. Sun \textit{et al.} \cite{sun2021time} developed a multi-objective optimization algorithm to simultaneously minimize the total transmission time, total performing time of UVAA, and total motion and hovering energy consumption of UAVs. Moreover, the authors in \cite{sun2022secure} studied a secure and energy-efficient relay communication problem, where the UVAA serves as an aerial relay for providing communications to the blocked or low-quality terrestrial networks. The authors designed an improved evolutionary computation method to jointly circumvent the effects of the known and unknown eavesdroppers and minimize the propulsion energy consumption of UAVs. 

\par Likewise, Sun \textit{et al}.~\cite{Sun2024} considered a UAV-enabled secure communication scenario, where multiple UAVs form a UVAA to perform CB for communicating with the remote BSs, and multiple known and unknown eavesdroppers exist to wiretap the information. Li \textit{et al}.~\cite{Li2024} studied the UAV-assisted IoT and introduced CB into IoTs and UAVs simultaneously to achieve energy and time-efficient data harvesting and dissemination from multiple IoT clusters to remote BSs. Huang \textit{et al}.~\cite{Huang2023} studied a dual-UAVs jamming-aided system to implement physical layer encryption in maritime wireless communication. Li \textit{et al}.~\cite{Li2022} considered a UAV-enabled data collection scenario, where CB is adopted to mitigate the interference of the LoS channel of UAV to the terrestrial network devices. Moorthy \textit{et al}.~\cite{KrishnaMoorthy2023} aimed at achieving the high data rate of CB of UAV swarm, and then designed a distributed solution algorithm based on a combination of echo state network learning and online reinforcement learning to maximize the throughput of UAV swarm. However, all of these abovementioned methods will take much time to compute the solution for a task. More seriously, when the task changes, the previous optimization solution is invalid, causing a longer latency for recalculation.

\par DRL has been an effective way to solve the online computation problems about UAV communications. For example, Wang \textit{et al.} \cite{wang2021trajectory} investigated a UAV-assisted IoT system, and they proposed a twin-delayed deep deterministic policy gradient (TD3)-based method that optimizes the trajectory of UAV to efficiently collect data from IoT ground nodes. Xie \textit{et al.} \cite{xie2021connectivity} proposed a multi-step dueling double DQN (multi-step D3QN)-based method to optimize the trajectory of cellular-connected UAV while guaranteeing the stable connectivity of UAV to the cellular network during its flight. Besides, MADRL can be also applied to solve the communication problems involving multiple UAVs. For instance, Zhang \textit{et al.} \cite{zhang2020uav} proposed a continuous action attention multi-agent deep deterministic policy gradient (CAA-MADDPG) algorithm to operate a UAV transmitter to communicate with ground users and multiple UAV jammers. Dai \textit{et al.} \cite{dai2022multi} studied the joint decoupled uplink-downlink association and trajectory design problem for full-duplex UAV network, with the purpose of maximizing the sum-rate of user equipments in both uplink and downlink. The authors proposed a MADRL-based approach and developed an improved clip and count-based proximal policy optimization (PPO) algorithm. However, in the above works, the cooperation pattern among UAVs is only the coordination, which is less complex than CB, and the completion of CB involves more UAVs. Hence, these abovementioned MADRL algorithms are not suitable for solving our formulated optimization problem. 

\par In summary, different from the above works, we investigate a joint problem about communication performance and energy consumption under a scenario where multiple UAVs perform CB. Then, the MADRL is adopted to efficiently solve our formulated optimization problem.

\begin{figure}[t]
	\centering
	\includegraphics[width=7.5cm]{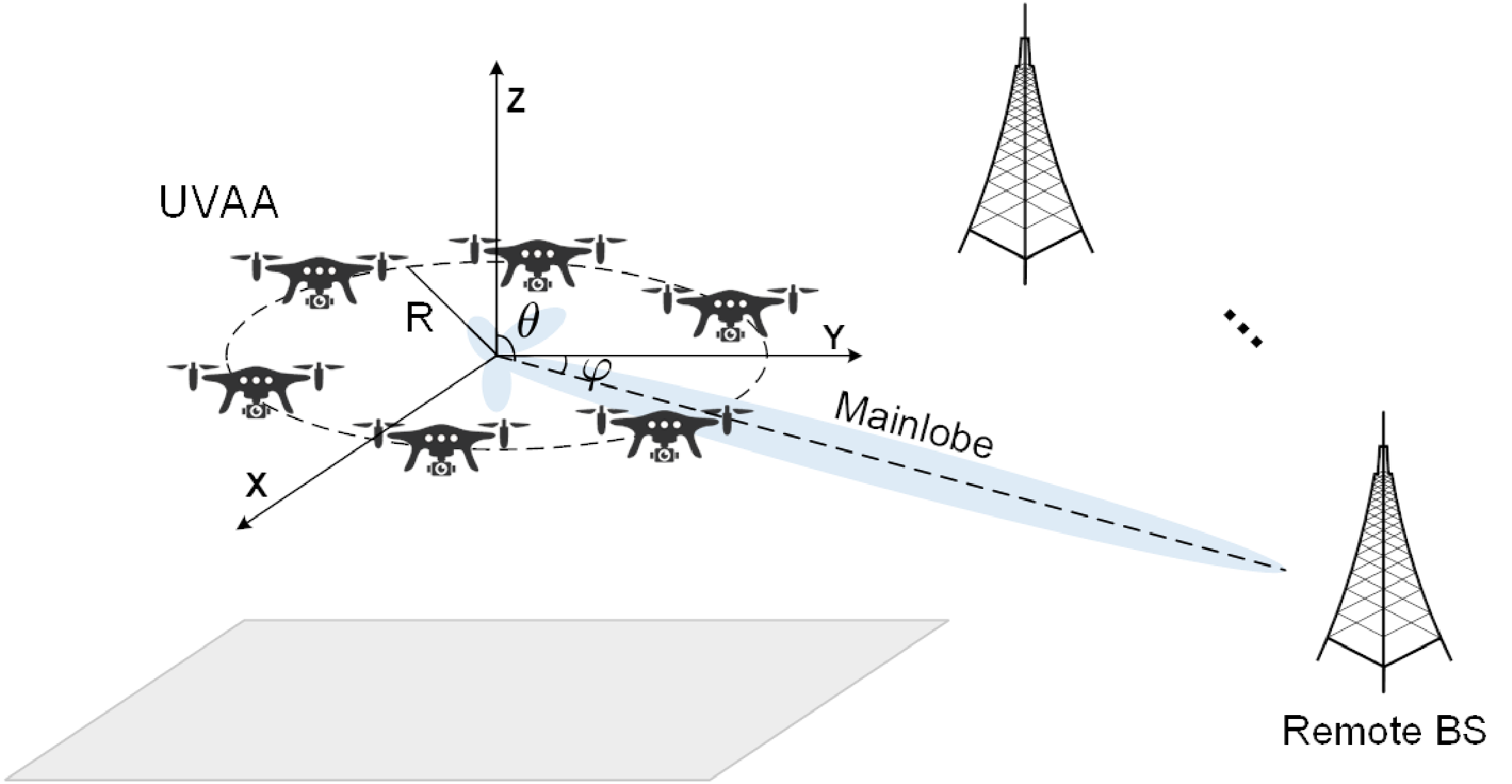}
	\caption{Sketch map of the considered UVAA communication system.}
	\label{fig:system_model}
\end{figure}

\section{System Model and Problem Formulation}
\label{sec:system_model_and_problem_formulation}

\par As shown in \figref{fig:system_model}, a UAV-assisted A2G wireless communication system is considered. Specifically, there are $N$ UAVs flying in a square area with length $L$, where the set of UAVs is denoted as $\mathcal{N} \triangleq \{i = 1, 2, \dots, N\}$. Some BSs are distributed in the far-field region, and their positions are assumed to be known by all UAVs. Each UAV is equipped with a single omni-directional antenna, and they will perform the communication tasks with these BSs. For example, their stored data or some generated emergency data needs to be uploaded to BSs. However, each UAV is not able to support such the long-distance communication due to the limited transmission ability, and they also can not fly to these BSs because of their own limited battery energy capacity. Note that for all UAVs, the distances among them are much shorter than those to BSs, which means that they are close to each other. Thus, all UAVs can fly closer to form an UVAA for performing CB to complete communication tasks. Moreover, the UVAA only communicates with one BS at each time as performing CB. 

\par In the considered system, the 3D Cartesian coordinate system is utilized. Specifically, the positions of the $i$-th UAV and the target BS are denoted as $(x_i, y_i, z_i)$ and $(x_{\mathrm{BS}}, y_{\mathrm{BS}}, 0)$, respectively. Furthermore, the elevation and azimuth angles from each UAV to the target BS can be computed by 
$\theta_{i} = \cos^{-1}(\frac{{0-z_{i}}}{{\sqrt{(x_{\mathrm{BS}}-x_{i})^{2} + (y_{\mathrm{BS}}-y_{i})^{2} + (0-z_{i})^{2}}}})$ and $\phi_{i} = \sin^{-1}(\frac{y_{\mathrm{BS}}-y_{i}}{\sqrt{(x_{\mathrm{BS}}-x_{i})^{2} + (y_{\mathrm{BS}}-y_{i})^{2}}})$, respectively.

\begin{table}[t]
	\renewcommand\arraystretch{1.05}	
	\caption{Summary of main notations}
	\centering
	\begin{tabular*}{\linewidth}{m{1.4cm}<{\raggedright} m{6.7cm}<{\raggedright}}
		\hline
		Notation & Definition \\  
		\hline
		\multicolumn{2}{c}{Notation used in system model} \\
		\hline
		$AF(\theta,\phi)$ & Array factor \\
		$B$ & Transmission bandwidth \\
		$d_{\mathrm{BS}}$ & Distance from origin of UVAA to target BS \\	
		$d_{min}$ & Minimal distance between two UAVs \\
		$E$ & Total energy consumption of UAV \\ 
		$E_{G}$ & Energy consumed by gravity during UAV flight \\ 
		$E_{K}$ & Energy consumption from kinetic energy during UAV flight \\ 
		$f_{c}$ & Carrier frequency \\ 
		$G(\cdot)$ & Array gain of UVAA \\
		$g_{c}$ & Channel power gain \\
		$H_{min}$ & Minimum flight altitude of UAV \\
		$H_{max}$ & Maximum flight altitude of UAV \\
		$I_{i}$ & Excitation current weight of the $i$-th UAV \\
		$L$ & Length of monitor area \\
		$m_{\mathrm{UAV}}$ & Mass of UAV \\ 
		$N$ & Number of UAVs \\
		$P_{\mathrm{LoS}}$ & LoS probability \\
		$P_{\mathrm{NLoS}}$ & NLoS probability \\
		$P_{t}$ & Total transmit power of UVAA \\
		$R(\cdot)$ & Transmission rate \\
		$w(\theta,\phi)$ & Magnitude of far-field beam pattern of UVA element \\
		$\alpha$ & Path loss exponent \\
		$\eta$ & Antenna array efficiency \\
		$\theta_{\mathrm{BS}}$ & Elevation angle between UVAA and target BS \\
		$\lambda$ & Wavelength \\
		$\mu_{\mathrm{LoS}}$ & Attenuation factor for LoS link \\
		$\mu_{\mathrm{NLoS}}$ & Attenuation factor for NLoS link \\
		$\sigma^{2}$ & Noise power \\
		$\phi_{\mathrm{BS}}$ & Azimuth angle between UVAA and target BS \\
		$\Psi_{i}$ & Initial phase of the $i$-th UAV \\
		\hline
		\multicolumn{2}{c}{Notation used in reinforcement learning} \\
		\hline
		$a_{i}$ & Action of the $i$-th agent \\
		$\boldsymbol{a}$ & Joint action of all agents\\
		$\mathcal{A}_{i}$ & action space of the $i$-th agent \\
		$A^{\boldsymbol{\pi}}(s^{t}, \boldsymbol{a}^{t})$ & Advantage function for joint policy $\boldsymbol{\pi}$\\
		$\mathrm{D^{max}_{KL}}(\cdot, \cdot)$ & Maximal KL-divergence between two policies\\
	    $\boldsymbol{g}^{k}_{i}$ & Gradient of maximization objective of the $i$-th agent\\
	    $\boldsymbol{H}^{k}_{i}$& Hessian of expected KL-divergence of the $i$-th agent \\
		$J(\boldsymbol{\pi})$ & Expected total reward for all agents \\
		$L^{\boldsymbol{\pi}}_{1:i}(\cdot, \cdot)$ & Surrogate equation of the $i$-th agent\\
		$o_{i}$ & Local observation of the $i$-th agent \\
		$\mathcal{O}_{i}$ & Observation space of the $i$-th agent\\
		$Q^{\boldsymbol{\pi}}(s^{t}, \boldsymbol{a}^{t})$ & State-action value function for joint policy $\boldsymbol{\pi}$ \\
		$r_{i}$ & Reward of the $i$-th agent \\
		$s$ & Global state\\
		$\mathcal{S}$ & Global state space \\
		$V^{\boldsymbol{\pi}}(s^{t})$ & State value function for joint policy $\boldsymbol{\pi}$ \\		
		$\alpha^{j}$ &  A coefficient is found by backtracking line search \\
		$\gamma$ & Discount factor\\
		$\delta$ &  KL constraint threshold\\
		$\rho_{\boldsymbol{\pi}}$ & Marginal state distribution for joint policy $\boldsymbol{\pi}$\\
		
		\hline
	\end{tabular*}
	
	\label{table:notation}
\end{table}

\subsection{UVAA Communication Model}
\par The goal of UVAA is to enable multiple UAVs to simulate the classical beamforming for achieving the far-field communication with remote BSs. The gain towards the target BS will be enhanced by the superposition of electromagnetic waves emitted from all UAVs through CB. Before the CB-enabled communication, all UAVs can be synchronized in terms of the carrier frequency, time and initial phase \cite{jayaprakasam2017distributed}, and then send the same data to the target BS.

\par Benefiting from the high altitude of UAV, a higher LoS probability between the UVAA and BS can be achieved. Additionally, the LoS probability depends also on the propagation environment and the statistic modeling of the building density. Hence, the LoS probability can be modeled as \cite{al2014optimal}
\begin{equation}\label{eq:LoS_probability}
	P_{\mathrm{LoS}} = \frac{1}{1+C\operatorname{exp}(-D[\theta_{\mathrm{rec}}-C])},
\end{equation}
where $C$ and $D$ are the parameters that depend on the propagation environment. $\theta_{\mathrm{rec}}$ is the elevation angle from the UVAA to the target BS. Then, the non-LoS (NLoS) probability is given by $P_{\mathrm{NLoS}}=1 - P_{\mathrm{LoS}}$. 

\par According to the probability of LoS and NLoS, the channel power gain can be expressed as 
\begin{equation}\label{eq:channel_power_gain}
	g_{c} = K^{-1}_{o}d^{-\alpha}_{\mathrm{BS}}[P_{\mathrm{LoS}}\mu_{\mathrm{LoS}} + P_{\mathrm{NLoS}}\mu_{\mathrm{NLoS}}]^{-1},
\end{equation}
where $K_{o}=(\frac{4 \pi f_{c}}{c})^{2}$, $d_{\mathrm{BS}}$ is the distance from the origin of UVAA to the target BS, $\mu_{\mathrm{LoS}}$ and $\mu_{\mathrm{NLoS}}$ are different attenuation factors for the LoS and NLoS links, respectively, 
$\alpha$ is a constant representing the path loss exponent, $f_{c}$ is the carrier frequency, and $c$ is the speed of light.

\par Combining the above channel model, the UVAA will perform CB to generate a beam pattern with a sharp mainlobe for the target BS.  
Then, the transmission rate can be expressed as \cite{mozaffari2018communications}
\begin{equation}\label{eq:transmission_rate}
	R(\boldsymbol{\mathrm{X}}, \boldsymbol{\mathrm{Y}}, \boldsymbol{\mathrm{Z}},  \boldsymbol{\mathrm{I}}) = B\log_2\left(1 + \frac{g_{c} P_{t} G(\boldsymbol{\mathrm{X}}, \boldsymbol{\mathrm{Y}}, \boldsymbol{\mathrm{Z}},  \boldsymbol{\mathrm{I}}) }{\sigma^2}\right),
\end{equation}
where $\boldsymbol{\mathrm{X}}=\{x_i\}^{N}_{i=1},
\boldsymbol{\mathrm{Y}}=\{y_{i}\}^{N}_{i=1}, \boldsymbol{\mathrm{Z}}= \{z_{i}\}^{N}_{i=1}, \boldsymbol{\mathrm{I}}=\{I_{i}\}^{N}_{i=1}$ represent the 3D coordinates and excitation current weights of all UAVs while communicating with the BS, $B$ is the transmission bandwidth, $P_t$ is the total transmit power of UVAA, and $\sigma^2$ is the noise power. In addition, $G(\boldsymbol{\mathrm{X}}, \boldsymbol{\mathrm{Y}}, \boldsymbol{\mathrm{Z}},  \boldsymbol{\mathrm{I}})$ is the array gain of UVAA toward the location of BS, which is given by \cite{mozaffari2018communications}
\begin{equation}\label{eq:array_gain}
	\begin{aligned}
		G(\boldsymbol{\mathrm{X}}, \boldsymbol{\mathrm{Y}}, \boldsymbol{\mathrm{Z}},  \boldsymbol{\mathrm{I}}) = \frac{4\pi|AF(\theta_{\mathrm{BS}}, \phi_{\mathrm{BS}})|^2w(\theta_{\mathrm{BS}}, \phi_{\mathrm{BS}})^2}{\displaystyle \int_{0}^{2\pi}\int_{0}^{\pi}|AF(\theta, \phi)|^2w(\theta, \phi)^2\sin{\theta}d\theta d\phi}\eta,
	\end{aligned}
\end{equation}
where $\eta \in [0, 1]$ is the antenna array efficiency, $(\theta_{\mathrm{BS}}, \phi_{\mathrm{BS}})$ represents the direction toward the target BS, and $w(\theta, \phi)$ is the magnitude of the far-field beam pattern of each UAV element. Note that the array gain is equal to the array directivity multiplied by $\eta$. The array directivity $D=U(\theta_{\mathrm{BS}}, \phi_{\mathrm{BS}})/U_{0}$, where $U(\theta_{\mathrm{BS}}, \phi_{\mathrm{BS}})=|AF(\theta_{\mathrm{BS}}, \phi_{\mathrm{BS}})|^2w(\theta_{\mathrm{BS}}, \phi_{\mathrm{BS}})^2$ is the maximum radiation intensity produced by the array factor (AF) to the BS, and $U_{0}=1/4\pi \int_{0}^{2\pi}\int_{0}^{\pi}|AF(\theta, \phi)|^2w(\theta, \phi)^2\sin{\theta}d\theta d\phi$ is the average radiation intensity. Besides, $w(\theta, \phi)$ is always equal to 0 dB in this work because the omni-directional antenna equipped on the UAV has the identical power constraints.

\par As for the AF, it is an important index to describe the status of beam pattern, which is written as \cite{garza2016design} \cite{sun2019energy} 
\begin{equation}\label{eq:array_factor}
	\begin{aligned}
		AF(\theta, \phi) = \sum_{i=1}^NI_{i}e^{j\Psi_{i}}e^{j[\frac{2\pi}{\lambda}(x_{i}\sin{\theta}\cos{\phi} + y_{i}\sin{\theta}\sin{\phi} + z_{i}\cos{\theta})]},
	\end{aligned}
\end{equation}
where $\theta\in[0, \pi]$ and $\phi\in[-\pi, \pi]$ are the elevation and azimuth angles, respectively. $\lambda$ is the wavelength and $\frac{2\pi}{\lambda}$ denotes the wave number. Additionally, $I_{i}$ and $\Psi_{i}$ are the excitation current weight and initial phase of the $i$-th UAV, where the phase synchronization is completed by compensating the distance between an UAV and the origin of UVAA \cite{jayaprakasam2017sidelobe}. Hence, $\Psi_{i}$ is expressed as
\begin{equation}\label{eq:initial_phase_of_i-th_UAV}
	\begin{aligned}
		&\Psi_{i} = \\&-\frac{2\pi}{\lambda}(x_{i}\sin{\theta_{\mathrm{BS}}}\cos{\phi_{\mathrm{BS}}} + y_{i}\sin{\theta_{\mathrm{BS}}}\sin{\phi_{\mathrm{BS}}} + z_{i}\cos{\theta_{\mathrm{BS}}}).
	\end{aligned}
\end{equation}
As can be seen, the locations and excitation current weights of UAVs are directly related to the AF. 

\subsection{UAV Energy Consumption Model}
\par In this work, we adopt a typical rotary-wing UAV. This is due to the fact that rotary-wing UAVs have the inherent benefit of hovering capabilities, which simplifies the difficulty of beam alignment and handling Doppler effects during transmission. In this work, the considered communication system is required to provide a stable and high-rate communication service for the far-field BSs. In this case, rotary-wing UAVs are more suitable for the considered scenario since they can hover for collecting and forwarding data. In general, the total energy consumption of a UAV includes two main components that are the propulsion and communication \cite{zeng2019energy}. Specifically, the propulsion energy consumption is related to the hovering and movement of the UAVs, and communication-related energy consumption is used for signal processing, circuits, transmitting and receiving, respectively. Typically, the propulsion energy consumption is much more than communication-related energy consumption (e.g., hundreds of watts (w) versus a few w) \cite{zeng2019energy} \cite{zeng2018trajectory}. Thus, the communication-related energy consumption is neglected in this work. Then, the propulsion power consumption of UAV with speed $v$ during the straight and horizontal flight is calculated as \cite{zeng2019energy}
\begin{equation}\label{eq:energy_comsumption_2D}
	\begin{aligned}
		P(v) = & P_B\left(1 + \frac{3v^2}{v^2_{tip}}\right) + P_I\left(\sqrt{1+\frac{v^4}{4v^4_0}} - \frac{v^2}{2v^2_0}\right)^{1/2} \\& + \frac{1}{2}d_0 \rho sAv^3,
	\end{aligned}
\end{equation}
where $P_{B}$ and $P_{I}$ are blade profile power and induced power under the hovering condition, respectively. $v_{tip}$ is the tip speed of rotor blade, and $v_0$ is mean rotor induced velocity in hovering. $d_0$, $\rho$, $s$, and $A$ denote the fuselage drag ratio, air density, rotor solidity and rotor disc area, respectively. Besides, the additional/less energy consumption caused by the acceleration/deceleration of horizontal flight is ignored because it only takes a small proportion of the total operation time of UAV manoeuvring duration \cite{zeng2019energy}.
\begin{figure}[t]
	\centering
	\hspace{-0.2cm}
	\includegraphics[width=7.5cm]{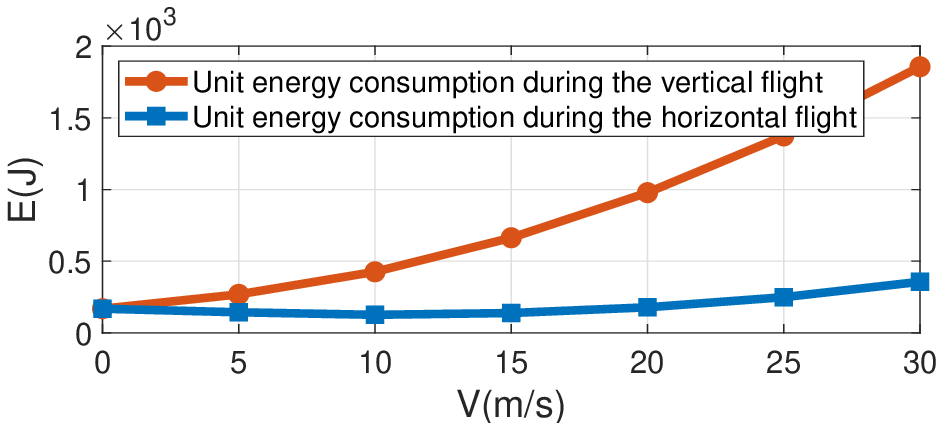}
	\caption{Energy consumption of UAV in the horizontal and vertical flights under different speeds.}
	\label{fig:E_vs_v}
\end{figure}
\par For arbitrary 3D UAV trajectory with UAV climbing and descending over time, the heuristic closed-form approximation of energy consumption is derived as \cite{zeng2019accessing}
\begin{equation}\label{eq:energy_comsumption_3D}
	\begin{aligned}
		E(T) \approx \int_0^{T} P(v(t))dt +\frac{1}{2}&m_{\mathrm{UAV}}(v(T)^2 - v(0)^2) \\
		&+m_{\mathrm{UAV}}g(h(T)-h(0)), 
	\end{aligned}
\end{equation}
where $T$ is the total flight time. $v(t)$ is the instantaneous UAV speed at time $t$. The second and third parts represent the kinetic energy consumption and potential energy consumption, respectively. $m_{\mathrm{UAV}}$ is the mass of UAV, and $g$ is the gravitational acceleration. As can be seen, \eqaref{eq:energy_comsumption_3D} has more terms than \eqaref{eq:energy_comsumption_2D}, meaning that more energy will be consumed in the vertical direction. Moreover, \figref{fig:E_vs_v} shows that the motion energy consumption of UAV per second in the vertical directioon is more than that in the horizontal direction under different speeds.
 
\par In this work, the UAV follows the flight rule that it first flies along the horizontal direction, then in the vertical direction, which is a common adopted flight strategy \cite{chou2019energy}\cite{yang2019energy}\cite{you2020hybrid}. Therefore, when a UAV reaches the target position, the corresponding energy consumption can be expressed as
\begin{equation}\label{eq:energy_comsumption_ind}
	\begin{aligned}
		E = P_H(v_H)t_H +  P_C(v_C)t_C +  P_D(v_D)t_D,
	\end{aligned}
\end{equation}
where $t_{H} = \Delta d/v_{H}$, $t_{C} = |\Delta h_{C}|/v_{C}$ and $t_{D} = |\Delta h_{D}|/v_{D}$. $\{P_H, P_C, P_D\}$ are the horizontal, climbing and descending flight powers, respectively, and $\{\Delta d, \Delta h_C, \Delta h_D\}$ are the corresponding flying distance during the horizontal, climbing and descending flights. Furthermore, $\{v_H, v_C, v_D\}$ represent the maximum-endurance (ME) speed of the horizontal, climbing and descending flights, respectively, which are the optimal UAV speed that maximizes the UAV endurance with any given onboard energy \cite{zeng2019accessing}.

\begin{remark}
    The challenge of adopting this type of rotary-wing UAVs lies in their weak transmission abilities and limited service time. To solve this issue, we correspondingly introduce the CB and give the problem formulation as follows.
\end{remark}

\subsection{Problem Formulation}
\par In the system, the UVAA is required to communicate with more than one BS before the battery runs out. To achieve the maximum transmission performance, all UAVs need to fly to better locations and adjust their excitation current weights. However, the motion energy consumption of UAVs will be increased during their movements, which will reduce the lifespan of UVAA. Therefore, the optimization problems about transmission rate and energy consumption should be considered simultaneously.

\par In this work, we aim to maximize the transmission rate of UVAA for communicating with the BS and minimize the motion energy consumption of all UAVs simultaneously, through optimizing the hovering positions and excitation current weights of all UAVs. Then, the UCBMOP can be formulated as 
\begin{subequations}
	\begin{align}
		&\mathop{\operatorname{max}}\limits_{\boldsymbol{\mathrm{X}}, \boldsymbol{\mathrm{Y}}, \boldsymbol{\mathrm{Z}},  \boldsymbol{\mathrm{I}}} & & f=\{R_T, -E_{total}\}\label{optimization_problem_1} \\
		&\quad~~\textrm{s.t.} & &C1: -L/2 \leq x_{i} \leq L/2, \forall i\in\mathcal{N} \label{optimization_problem_2},\\
		&&&C2: -L/2 \leq y_{i} \leq L/2, \forall i\in\mathcal{N} \label{optimization_problem_3},\\
		&&&C3:  H_{min} \leq z_{i} \leq H_{max}, \forall i\in\mathcal{N} \label{optimization_problem_4},\\
		&&&C4:  0\leq I_{i} \leq1, \forall i\in\mathcal{N}  \label{optimization_problem_5},\\
		&&&C5:  d_{i, j} \geq d_{min}, \forall i, j\in \mathcal{N}, \label{optimization_problem_6}
	\end{align} 
	\label{optimization_problem_6}
\end{subequations}
\noindent where $R_{T}$ is the transmission rate of UVAA to a specific BS defined by \eqaref{eq:transmission_rate}. The motion energy consumption of all UAVs is $E_{total} = \sum_{i=1}^N E_{i}$, where $E_{i}$ is the motion energy consumption of UAV $i$ that can be obtained according to the \eqaref{eq:energy_comsumption_ind}. Constraints C1, C2 and C3 together specify the flight area of UAV, where $L$ is the length of monitor area, and $H_{min}$ and $H_{max}$ are altitude constraints of the UAVs. Constraint C4 specifies the adjustment range of excitation current weight. Moreover, Constraint C5 requires any two adjacent UAVs to keep a certain distance to avoid the collision, where $d_{min}$ is set as the minimum distance.
  
\section{MADRL-Based UAV-enabled Collaborative Beamforming}
\label{sec:HATRPO_UCB}

\par The formulated UCBMOP is challenging and cannot be solved by traditional optimization methods, and the reasons are as follows:

\begin{itemize}

    \item \textit{UCBMOP is a non-convex and long-term optimization problem}. Specifically, due to the complex non-linear relationship between UAV coordinates and excitation current weights with antenna gain, the optimization of the virtual antenna array transmission performance is a non-convex optimization problem. Meanwhile, UCBMOP involves long-term optimization objectives and needs an effective algorithm that can achieve long-term optimal solutions. As such, UCBMOP is challenging and cannot be solved by traditional convex optimization.

    \item \textit{UCBMOP is an NP-hard problem}. Specifically, UCBMOP can be simplified as a typical nonlinear multidimensional 0-1 knapsack problem that has been demonstrated to be NP-hard~\cite{Goos2020}. Therefore, the proposed UCBMOP is also NP-hard. Such NP-hard problems are challenging for traditional optimization methods to solve in polynomial time.

    \item \textit{UCBMOP is a complex large-scale optimization problem}. Specifically, the considered scenarios often involve a large number of UAVs, while each UAV also has multiple variables to be optimized (i.e., 3D coordinates and excitation current weights). In addition, the completion of CB also needs the complex cooperation of all UAVs. Hence, the complexity and massive dimension of decision variables make the traditional optimization method unable to solve it efficiently.

    \item \textit{UCBMOP operates within a dynamic environment}. In particular, UAVs in these scenarios must make decisions to adapt to highly dynamic channel conditions and the unpredictability of natural surroundings. Additionally, real-time response is crucial for UAVs, as prolonged hovering in the air results in significant energy consumption. Consequently, conventional offline algorithms or approaches reliant on temporary calculations cannot solve this problem.
    
\end{itemize}

\par Different from the traditional optimization methods, deep reinforcement learning can have real-time responses and handle the complexity and large-scale decision variables of UCBMOP. However, in the considered system, UAVs need to collaborate to achieve transmission gain while simultaneously competing to minimize their individual energy consumption. Such tasks are distributed and require multiple UAVs to work together to achieve common objectives. Standard reinforcement learning assumes individual agents to be isolated, making it less suitable for the scenario involving multiple decision-makers.

\begin{remark}
    In this case, MADRL has the capability to decentralize decision-making and adaptability for the changing environments, and enabling effective responses to complex interactions among multiple agents~\cite{Feriani2021}. Moreover, MADRL allows agents to access the global state of the UAV collaborative beamforming system, providing them with information about the entire system and reducing non-stationarity. In addition, the scalability and robustness of MADRL to partial observability also enhance its applicability in the considered dynamic and distributed scenarios. Thus, we propose an MADRL-based method that is able to solve the formulated UCBMOP by learning the strategy in the considered dynamic scenario. 
\end{remark}

 
\subsection{Markov Game for UCBMOP}
\par In this section, the formulated UCBMOP is transformed into a Markov game \cite{littman1994markov} that is defined as $\langle \mathcal{N}, \mathcal{S}, \{\mathcal{O}_{i}\}_{i\in \mathcal{N}}, \{\mathcal{A}_{i}\}_{i\in \mathcal{N}}, P, \{R_{i}\}_{i\in \mathcal{N}},  \gamma \rangle$. At time slot $t$, all agents are with state $s^{t}\in \mathcal{S}$, each agent uses its policy $\pi_{\theta_{i}}$ to take an action $a^{t}_{i}\in \mathcal{A}_{i}$ according to its local observation $o^{t}_{i}\in \mathcal{O}_{i}$, and it will receive a reward $r^{t}_{i}=R_{i}(s^{t}, \boldsymbol{a}^{t})$ from the environment, where $\boldsymbol{a}^{t}=\{a^{t}_{1},\dots,a^{t}_{N}\}$ denotes the joint action. Then, all agents move to the next state $s^{t+1}$ with probability $P(s^{t+1}|s^{t}, \boldsymbol{a}^{t})$ and each of them receives its local observation $o^{t+1}_{i}$ correlated with the state. All agents aim to maximize their own expected total reward $J(\pi_{\theta_{i}})=\mathbb{E}_{s^{0},\boldsymbol{a}^{0},\dots}[\sum^{\infty}_{t=0}\gamma^{t} r^{t}_{i}]$. The details about the state and observation space, the action space and the reward function are introduced as follows.

\subsubsection{State and Observation Space}
\par At each time slot, every agent observes its own local state information to make the decision. In the system, the local observation of any agents (i.e., UAVs) includes three parts that are its own related information, the relevant information of other UAVs and the related information of BS. Specifically, the local observation of the $i$-th agent can be expressed as
\begin{equation}\label{observation_space}
	\begin{aligned}
		o_{i} &= \{\theta_i, \phi_i, d_{i}, d_{i,o}, I_{i},  \{d_{j,i}, I_{j}\}_{j\in\mathcal{N}, j\neq i}, x^{\prime},y^{\prime},z^{\prime}\},
	\end{aligned}
\end{equation}
\noindent where $(\theta_{i}, \phi_{i}, d_{i})$ is the spherical coordinate of the target BS relative to the $i$-th UAV. However, when the UVAA will communicate with more than one BS, then the gap among different BSs regarding $d_{i}$ is very large. Furthermore, all UAVs only fly in the fixed area. To eliminate the gap and let the UAV make decisions more efficiently, we compute a reference point in the monitor area that is closest to the target BS, where the coordinate of reference point is denoted as $(x^{\prime}, y^{\prime}, z^{\prime})$. Then, the distance between the $i$-th UAV and the reference point is regarded as $d_{i}$. Moreover, $d_{i,o}$ is the distance between the $i$-th UAV and the origin of UVAA, and $I_{i}$ is the excitation current weight of the $i$-th UAV. The information of other UAVs contains their own excitation current weights and the distances to the $i$-th UAV. As for the information about the target BS, its coordinate is replaced by the reference point, which will be beneficial for the agent to learn the better strategy.

\par As for the global state of the $i$-th agent $s_{i}$, instead of using the concatenation of local observations or the environment-provided global state, we propose an agent-specific global state that combines some agent-related features (e.g., $(\theta_{i}, \phi_{i}, d_{i})$, $d_{i,o}$, $d_{j,i}$) and the global information provided from the environment which includes the coordinates and excitation current weights of all UAVs, as well as the coordinate of reference point, which is inspired from \cite{yu2021surprising}.

\begin{figure}[t]
	\centering
	\includegraphics[width=7.5cm]{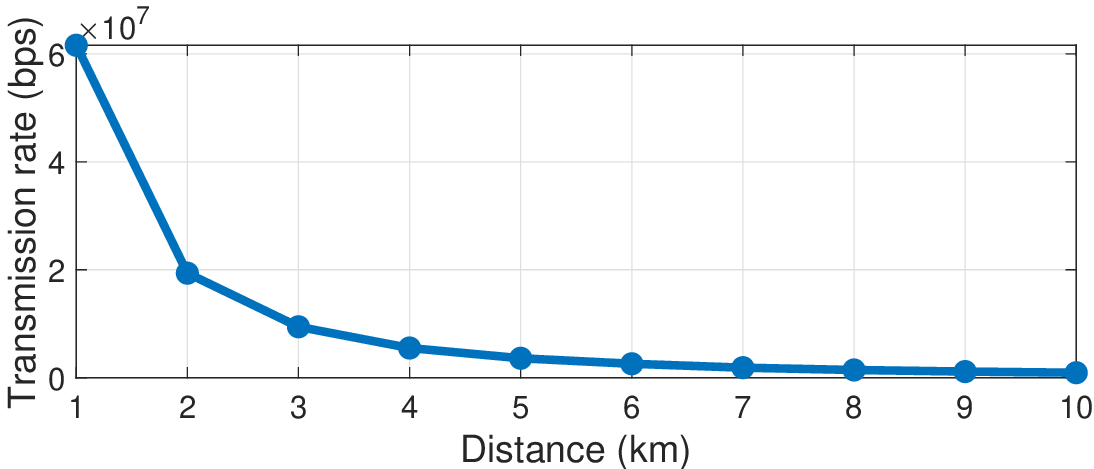}
	\caption{The transmission rate of UVAA to BSs at different distances when performing CB communication. The parameters about LoS probability $C$ and $D$ are set as 10 and 0.6, respectively.}
	\label{fig:R}
\end{figure}

\subsubsection{Action Space}
\par According to the local observation of UAVs, each UAV will take an action to fly to a certain location with the optimized excitation current weight for performing CB. Then, the action of each agent can be defined by 
\begin{equation}\label{action_space}
	\begin{aligned}
		a_{i} = \{x_i, y_i, z_i, I_i\},
	\end{aligned}
\end{equation}
where $(x_i, y_i, z_i)$ is the location to which the $i$-th UAV is going to fly, and $I_i$ is the excitation current weight of the $i$-th UAV in the next time slot.

\subsubsection{Reward Function}

\par The reward function is used to evaluate how well an agent executes the actions. Specifically, a suitable reward function can make the agent learn the best strategy. UCBMOP involves two optimization objectives, which are maximizing the transmission rate of UVAA for communicating with the BS, i.e., $R_T$, and minimizing the motion energy consumption of all UAVs, i.e., $E_{total}$. To this end, we design a reward function by considering the following aspects:

\begin{enumerate}
    \item \textit{Reward Function for Optimizing $R_T$}. We find that there are three key components that affect the transmission rate $R_T$, which are the transmit power and gain of UVAA, transmission distance, and A2G communication angle. Specifically, the transmit power and gain of UVAA determine the transmission performance of UVAA, the transmission distance indicates the signal fading degree, and the air-to-ground communication angle affects the A2G channel condition. Note that we do not directly set the transmission rate as a reward since it changes significantly with the transmission distance, making the DRL method unstable and decreasing the applicability of the model.
    
    \par According to the analysis above, \textit{first}, we regard the multiplication of the array gain and the total transmit powers as a reward, i.e., $r^\mathrm{TR} = GP_t$. Note that this reward is shared by all agents. \textit{Then}, we set the distance between the UAV and BS as a penalty for each agent. However, the UAV only flies within a fixed area, and then the aforementioned distance is replaced by the one between the UAV and the reference point of target BS, which is denoted as $r^\mathrm{U2B}_i = -d_{i,r}$. \textit{Finally}, as shown in Eq.~\eqref{eq:LoS_probability}, the LoS probability is determined mainly by the height of the UAVs. Therefore, we set a reward to optimize the flight altitude of each UAV, denoted as $r_i^\mathrm{H} = z_i$. As such, the designed reward function considers all controllable components of the transmission rate, thereby achieving the optimization of the first objective. 

    \item \textit{Reward Function for Optimizing $E_{total}$}. To achieve this goal, we regard the energy consumption of each UAV as a penalty, which is denoted as $r_i^\mathrm{EC} = -E_i$. As such, the optimization of the second objective can be achieved by the reward function.

    \item \textit{Reward Function for Coordinating $R_T$ and $E_{total}$}. To this end, we aim to enhance the connectivity among UAVs. Compared with $R_T$, the optimization of $E_{total}$ is more directly tied to the action space (i.e., reducing the movement of the UAVs), which may lead the algorithm to prioritize the optimization of $E_{total}$ and under-optimize $R_T$. To overcome this issue, we try to concentrate the array elements (i.e., the UAVs in UVAA) on an applicable distance to achieve a higher CB performance~\cite{Salama2019}~\cite{sun2022secure}, thereby facilitating the optimization of $R_T$. Specifically, we design a reward to coordinate the UAVs by minimizing the cumulative distances between the current UAV and other UAVs, as well as between the current UAV and the origin of the UVAA, i.e., 
\begin{equation}\label{U2U_reward}
	\begin{aligned}
		r^{\mathrm{U2U}}_i = -\frac{1}{\kappa}(\sum_{j \in \mathcal{N}, j \neq i} d_{j,i} + d_{i, o}),
	\end{aligned}
\end{equation}
\noindent where $\kappa$ is a large constant for scaling. Since this reward decreases the difficulty of optimizing $R_T$, the designed reward function can well coordinate the optimization processes of the two optimization objectives.
\end{enumerate}

\par In summary, the whole reward function of each agent is 
\begin{equation}\label{the_whole_reward_function}
	\begin{aligned}
		r_i = \omega_{1}r^{\mathrm{TR}} + \omega_{2}r^{\mathrm{H}}_{i} + \omega_{3}r^{\mathrm{EC}}_{i} + \omega_{4}r^{\mathrm{U2B}}_{i} + \omega_{5}r^{\mathrm{U2U}}_{i},
	\end{aligned}
\end{equation}
\noindent where $\omega_{1}$, $\omega_{2}$, $\omega_{3}$, $\omega_{4}$, and $\omega_{5}$ represent weight coefficients of different parts. Moreover, it is also worth noting that when any two UAVs violate the minimum separation constraint, they will receive a negative reward. These weighting coefficient values can be determined by considering their value ranges, importance, and characteristics. Specifically, larger weights are assigned to rewards with wider ranges to appropriately reflect their significance in optimization. Moreover, higher weights are allocated to rewards that play a more critical role in achieving the optimization objectives. In addition, the weights can also be adjusted based on the unique characteristics of each reward, ensuring that the optimization process addresses their specific complexities.

\begin{figure*}[t]
	\centering
	\includegraphics[width=15.5cm]{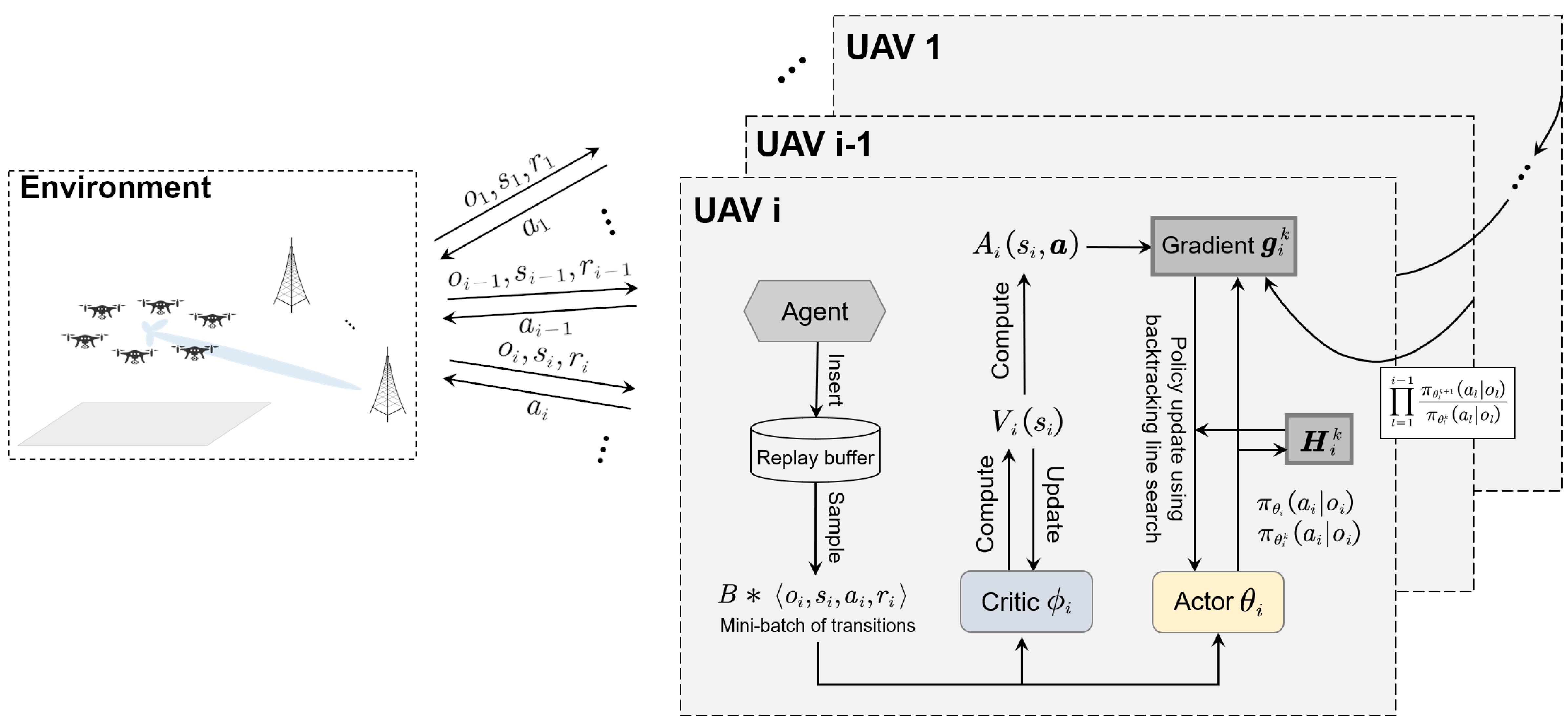}
	\caption{Flowchart of HATRPO-UCB algorithm for UAV-enabled CB.}
	\label{fig:framework_of_HATRPO_UCB}
\end{figure*}

\subsection{HATRPO-UCB Algorithm}

\par In this section, we propose a HATRPO-UCB, which extends from conventional HATRPO to find the solution strategy for UCBMOP. Specifically, each UAV acts as an agent that has an actor network and a critic network consisting of DNNs. The algorithm uses the sequential policy update scheme to train all agents. The scheme can make the update of the current agent includes the updates of previous agents, and thus our algorithm can make the actor networks of all agents learn the joint policy. Moreover, each agent has not only the individual rewards (i.e., $r_i^\mathrm{H}$, $r_i^\mathrm{EC}$, etc.) but also the shared rewards (i.e., $r^\mathrm{TR}$). Meanwhile, we design an agent-specific global state that contains global information provided by the environment and some features from the local observation. As such, each UAV is guided by the shared rewards, and considers the global information and local observation, thereby obtaining a policy associated to the global state. We begin with introducing the conventional HATRPO.


\subsubsection{Conventional HATRPO}
\par HATRPO was proposed by Kuba \textit{et al.} \cite{kuba2021trust}, which has the better performance than other baseline MADRL algorithms in many multi-agent tasks (e.g., Multi-Agent MuJoCo \cite{de2020deep}, StarCraftII Multi-Agent Challenge (SMAC) \cite{samvelyan2019starcraft}). HATRPO applies the trust region learning \cite{10.5555/645531.656005} to MADRL successfully, achieving the monotonic improvement guarantee for joint policy of multiple agents such as the trust region policy optimization (TRPO) algorithm \cite{schulman2015trust}. The detailed theory of HATRPO is introduced in the following.

\par In the beginning, a cooperative Markov game with $N$ agents exists. A joint policy $\boldsymbol{\pi}=(\pi_{1},\dots,\pi_{N})$ to all agents exists. At time slot $t$, the agents are at state $s^{t}$. Each agent utilizes its own policy $\pi_{i}$ to take an action $a^{t}_{i}$, and then the actions of all agents form the joint action $\boldsymbol{a}^{t}=(a^{t}_{1},\dots,a^{t}_{N})$. Afterwards, all agents receive a joint reward $r^{t}$, and move to a new state $s^{t+1}$ with probability $P(s^{t+1}|s^{t}, \boldsymbol{a}^{t})$. The goal of all agents is to maximize the expected total reward:
\begin{equation}\label{eq:expected_total_reward}
	\begin{aligned}
		J(\boldsymbol{\pi})=\mathbb{E}_{s^{0:\infty}\sim\rho^{0:\infty}_{\boldsymbol{\pi}},\boldsymbol{a}^{0:\infty}\sim\boldsymbol{\pi}}\left[\sum^{\infty}_{t=0}\gamma^{t}r^{t}\right],
	\end{aligned}
\end{equation}
where $\gamma\in [0, 1)$ is the discount factor and $\rho_{\boldsymbol{\pi}}$ is the marginal state distribution \cite{kuba2021trust}. Besides, the state value function and the state-action value function are defined as 
$V^{\boldsymbol{\pi}}(s^{t}) = \mathbb{E}_{\boldsymbol{a}^{t:\infty}\sim \boldsymbol{\pi}, s^{t+1:\infty}\sim P}\left[\sum_{l=0}^\infty \gamma^l r^{t+l}\right]$ and $Q^{\boldsymbol{\pi}}(s^{t}, \boldsymbol{a}^{t}) = \mathbb{E}_{s^{t+1:\infty}\sim P, \boldsymbol{a}^{t+1:\infty}\sim \boldsymbol{\pi}}\left[\sum_{l=0}^\infty \gamma^l r^{t+l}\right]$. The advantage function is written as $A^{\boldsymbol{\pi}}(s^{t}, \boldsymbol{a}^{t}) = Q^{\boldsymbol{\pi}}(s^{t}, \boldsymbol{a}^{t}) - V^{\boldsymbol{\pi}}(s^{t})$.

\par To extend the key idea of TRPO to MADRL, a sequential policy update scheme \cite{kuba2021trust} is introduced. Then, the following inequality can be derived to achieve the monotonic improvement guarantee for joint policy.
\begin{equation}\label{eq:inequal_equation_madrl}
	\begin{aligned}
		J(\boldsymbol{\bar{\pi}}) \geq J(\boldsymbol{\pi}) + \sum^N_{i=1}[L^{\boldsymbol{\pi}}_{1:i}(\boldsymbol{\bar{\pi}}_{1:i-1}, \bar{\pi}_i) - C\mathrm{D^{max}_{KL}}(\pi_{i}, \bar{\pi}_{i})].
	\end{aligned}
\end{equation} 
\noindent where $\boldsymbol{\bar{\pi}}$ is the next candidate joint policy. $\mathrm{D^{max}_{KL}}(\pi_{i}, \bar{\pi}_{i})= \mathrm{max}_{s}\mathrm{D_{KL}}(\pi_{i}(\cdot|s), \bar{\pi}(\cdot|s)) $ is the maximal KL-divergence to measure the gap between two policies. $C=\frac{4\gamma \mathrm{max}_{s,\boldsymbol{a}}|A^{\boldsymbol{\pi}}(s, \boldsymbol{a})|}{(1-\gamma)^2}$ is the penalty coefficient. Moreover,  $L^{\boldsymbol{\pi}}_{1:i}(\boldsymbol{\bar{\pi}}_{1:i-1}, \bar{\pi}_i)$ is a surrogate equation of the $i$-th agent, which is expressed as
\begin{equation}\label{surrogate_equation}
	\begin{aligned}
		L^{\boldsymbol{\pi}}_{1:i}(&\boldsymbol{\bar{\pi}}_{1:i-1}, \bar{\pi}_{i})\\
		&=\mathbb{E}_{s\sim \rho_{\boldsymbol{\pi}},\boldsymbol{a}_{1:i-1}\sim \boldsymbol{\bar{\pi}}_{1:i-1}, a_{i}\sim \bar{\pi}_{i}}[A^{\boldsymbol{\pi}}_{i}(s, \boldsymbol{a}_{1:i-1}, a_{i})],
	\end{aligned}
\end{equation}
where $A^{\boldsymbol{\pi}}_{i}(s, \boldsymbol{a}_{1:i-1}, a_{i})$ is the local advantage function of each agent. Based on the above theory, all agents can update their policies sequentially, and the updating procedure is given as
\begin{equation}\label{eq:policy_update_of_HATRPO}
	\begin{aligned}
		\pi^{k+1}_{i} = \underset{\pi_{i}}{\operatorname{arg\, max}}(L^{\boldsymbol{\pi}^{k}}_{1:i}(\boldsymbol{\pi}^{k+1}_{1:i-1}, \pi_{i}) - C\mathrm{D^{max}_{KL}}(\pi^{k}_{i}, \pi_{i})).
	\end{aligned}
\end{equation}
Furthermore, the sequential updating scheme does not require that the updating order of all agents is fixed, meaning that the updating order can be flexibly adjusted at each iteration. As such, the algorithm can adapt to dynamic changes in the environment and progressively assimilate new information in non-static conditions.

\par To implement the above procedure for parameterized joint policy $\boldsymbol{\theta}=(\theta_{1},\dots,\theta_{N})$ in practice, in HATRPO, the amendment similar to TRPO is performed. As the maximal KL-divergence penalty $\mathrm{D^{max}_{KL}}(\pi_{\theta^{k}_{i}}, \pi_{\theta_{i}})$ is hard to compute, it is replaced by the expected KL-divergence constraint $\mathbb{E}_{s\sim \rho_{\boldsymbol{\pi}_{\boldsymbol{\theta}^{k}}}}[\mathrm{D_{KL}}(\pi_{\theta^{k}_{i}}(\cdot|s), \pi_{\theta_{i}}(\cdot|s))]\leq \delta$, where $\delta$ is a threshold hyperparameter. Then, at the $k$+$1$-th iteration, given a permutation, all agents can optimize their policy parameters sequentially according to the method as follows: 
\begin{equation}\label{eq:practical_policy_update_of_HATRPO}
	\begin{aligned}
		&\theta^{k+1}_{i}=\\
		&\underset{\theta_{i}}{\operatorname{arg\, max}}\mathbb{E}_{s\sim \rho_{\boldsymbol{\pi}_{\boldsymbol{\theta}^{k}}},\boldsymbol{a}_{1:i-1}\sim \boldsymbol{\pi}_{\boldsymbol{\theta}^{k+1}_{1:i-1}}, a_{i}\sim \pi_{\theta_{i}}}[A^{\boldsymbol{\pi}_{\boldsymbol{\theta}^{k}}}_{i}(s, \boldsymbol{a}_{1:i-1}, a_{i})], \\
		&\operatorname{s.t.}  \mathbb{E}_{s\sim \rho_{\boldsymbol{\pi}_{\boldsymbol{\theta}^{k}}}}[\mathrm{D_{KL}}(\pi_{\theta^{k}_{i}}(\cdot|s), \pi_{\theta_{i}}(\cdot|s))]\leq \delta.
	\end{aligned}
\end{equation}
For the computation of the above equation, similar to TRPO, the linear approximation to the objective function and the quadratic approximation to KL constraint are applied, derivating a closed-form updating scheme that is shown as
\begin{equation}\label{eq:closed_form_policy_update_of_HATRPO}
	\begin{aligned}
		\theta^{k+1}_{i}=\theta^{k}_{i}+\alpha^{j}\sqrt{\frac{2\delta}{\boldsymbol{g}^{k}_{i}(\boldsymbol{H}^{k}_{i})^{-1}\boldsymbol{g}^{k}_{i}}}(\boldsymbol{H}^{k}_{i})^{-1}\boldsymbol{g}^{k}_{i}.
	\end{aligned}		
\end{equation}
$\alpha^{j}\in (0, 1)$ is a coefficient that is found via backtracking line search. $\boldsymbol{H}^{k}_{i}=\nabla^{2}_{\theta_{i}}\mathbb{E}_{s\sim \rho_{\boldsymbol{\pi}_{\boldsymbol{\theta}^{k}}}}[\mathrm{D_{KL}}(\pi_{\theta^{k}_{i}}(\cdot|s), \pi_{\theta_{i}}(\cdot|s))]|_{\theta_{i}=\theta^{k}_{i}}$ is the Hessian of the expected KL-divergence. $\boldsymbol{g}^{k}_{i}$ is the gradient of the objective in \eqaref{eq:practical_policy_update_of_HATRPO}, whose computation requires the estimation about $\mathbb{E}_{\boldsymbol{a}_{1:i-1}\sim \boldsymbol{\pi}_{\boldsymbol{\theta}^{k+1}_{1:i-1}}, a_{i}\sim \pi_{\theta_{i}}}[A^{\boldsymbol{\pi}_{\boldsymbol{\theta}^{k}}}_{i}(s, \boldsymbol{a}_{1:i-1}, a_{i})]$. According to the derivation in \cite{kuba2021trust}, 
given a batch $B$ of trajectories with length $T$, $\boldsymbol{g}^{k}_{i}$ can be computed as
\begin{equation}\label{eq:the_computation_of_gradient}
	\begin{aligned}
		\boldsymbol{g}^{k}_{i}=\frac{1}{B}\sum\limits^{B}_{b=1}\sum\limits^{T}_{t=0}M_{1:i}(s^{t},\boldsymbol{a}^{t})\nabla_{\theta_{i}}\operatorname{log}\pi_{\theta_{i}}(a^{t}_{i}|s^{t})|_{\theta_{i}=\theta^{k}_{i}},
	\end{aligned}
\end{equation} 
where
\begin{equation}\label{eq:M}
	\begin{aligned}
		M_{1:i}(s^{t},\boldsymbol{a}^{t})=\frac{\boldsymbol{\pi}_{\boldsymbol{\theta}^{k+1}_{1:i-1}}(\boldsymbol{a}^{t}_{1:i-1}|s^{t})}{\boldsymbol{\pi}_{\boldsymbol{\theta}^{k}_{1:i-1}}(\boldsymbol{a}^{t}_{1:i-1}|s^{t})}A^{\boldsymbol{\pi}_{\boldsymbol{\theta}^{k}}}(s^{t}, \boldsymbol{a}^{t}).
	\end{aligned}
\end{equation} 
$\frac{\boldsymbol{\pi}_{\boldsymbol{\theta}^{k+1}_{1:i-1}}(\boldsymbol{a}^{t}_{1:i-1}|s^{t})}{\boldsymbol{\pi}_{\boldsymbol{\theta}^{k}_{1:i-1}}(\boldsymbol{a}^{t}_{1:i-1}|s^{t})}$ is the compound policy ratio about the previous agents $1:i-1$, which can be computed easily after these agents complete their updates.

\par Although HATRPO has the excellent performance, it still faces many challenges in solving the UCBMOP. \emph{\textbf{First}}, the formulated problem involves the complex cooperation among UAVs. Furthermore, only one time slot exists for each episode in the problem, which is different from other traditional MADRL tasks. Thus, it is necessary to take measures to connect the algorithm with the problem closely. \emph{\textbf{Second}}, there are many UAVs in the problem, which may increase the input dimension of critic network substantially, influencing the critic learning. \emph{\textbf{Finally}}, the conventional HATRPO uses the Gaussian distribution to sample actions. However, all actions have the finite range in the problem. Thus, using the Gaussian distribution may induce the bias to the policy gradient estimation. Accordingly, these reasons motivate us to propose the HATRPO-UCB, and the details are described in the following sections.

\subsubsection{Algorithm Design of HATRPO-UCB}
\par In the proposed HATRPO-UCB, we take conventional HATRPO as the foundation algorithm framework, and propose three techniques that are \emph{observation enhancement}, \emph{agent-specific global state} and \emph{Beta distribution for policy}, to enhance the performance of the algorithm.

\par \emph{\textbf{Observation enhancement:}} In MADRL, the environmental information is usually the state that guides the agents to take action. However, in UCBMOP, each episode only has one time slot, and the UVAA may communicate with different remote BSs at various episodes. In this case, the state consisting of the simple Cartesian coordinates of the UAVs and BSs may not precisely characterize their positional relationship. This is because the distance between the UVAA and BSs is significantly larger than the relative distance between the UAVs, and even the same set of UAV coordinates may represent different environmental states when serving various BSs (e.g., transmission angle and distance). To overcome this issue, we propose two observation enhancement strategies as follows. \textit{First}, we design a spherical coordinate enhancement strategy with dynamically changing origins to represent the UAV positions. Instead of the Cartesian coordinate of each UAV, we set the state to the spherical coordinates with the BS in the current episode as the origin. In this way, the state can adequately express the information about the orientation and distance between the BS and the UVAA. \textit{Second}, we propose a position representation based on reference points to characterize the BS locations. Specifically, we regard the closest point to the target BS in the monitor area as the reference point and then adopt this reference point to replace the target BS as the state. This adjustment retains the BS information but amplifies the position differences between the UAVs, which better expresses the relative positions of the UAVs. The enhanced state and observation contain representative and adaptive information about the environment, which ensures the proposed method remains effective when the location of the served BS at different episodes changes significantly.

\par \emph{\textbf{Agent-specific global state:}} In general, the global state employed by MADRL algorithms has two forms, which are the concatenation of all local observations and environment-provided global state. However, in the considered scenarios, the number of agents (i.e., UAVs) is often large, which means that the input dimension of the critic network grows a lot when using the concatenation of all local observations. This condition will increase the learning burden, thereby making the performance of critic learning degrade.  Additionally, the environment-provided global state is also not an appropriate choice because its contained information is not sufficient, which only contains the coordinates and excitation current weights of all UAVs, as well as the coordinates of the reference point. Inspired by \cite{yu2021surprising}, we design an agent-specific global state that combines the global information provided by the environment and some features from local observation, such as the spherical coordinate of the target BS that is relative to the UAV $(\theta_{i}, \phi_{i}, d_{i})$ and the distance between two UAVs $d_{j,i}$, as the input of the critic network. This agent-specific global state can improve the fitting performance of critic learning and thus achieve better MADRL performance.
 
\par \emph{\textbf{Beta distribution for policy:}} In the conventional HATRPO, the actor uses the Gaussian distribution to sample actions. However, the Gaussian distribution has been demonstrated to have the negative impact on reinforcement learning \cite{chou2017improving}. Specifically, for an action with the finite interval, the Gaussian distribution with infinite support will define the action range out of its boundary. Furthermore, the probability density of all actions is greater than 0, but the probability of actions beyond boundary should equal to 0. Hence, in our scenario where all actions of UAV have the finite interval, it is inevitable that some actions sampled from the Gaussian distribution will be truncated, leading to the boundary effect. Then, the boundary effect may cause bias about policy gradient estimation.

\par To handle the above issue, the authors of \cite{chou2017improving} conducted the research on Beta distribution for reinforcement learning. The study indicates that the Beta distribution does not have a bias because the Beta distribution has a finite range (i.e., [0, 1]), and thus the probability density of actions beyond the boundary is guaranteed to be 0. Furthermore, experimental results show that the Beta distribution can make reinforcement learning algorithms converge faster and obtain higher rewards. Thus, in this paper, we choose the Beta distribution instead of the original Gaussian distribution, and then the Beta distribution is defined as
\begin{equation}\label{eq:Beta_distribution}
	\begin{aligned}
		f(x;\alpha,\beta)=\frac{\Gamma(\alpha+\beta)}{\Gamma(\alpha)\Gamma(\beta)}x^{\alpha-1}(1-x)^{\beta-1},
	\end{aligned}
\end{equation}
where $\alpha$ and $\beta$ are the shape parameters, and $\Gamma(\cdot)$ is the $\emph{Gamma}$ function that extends factorial to real numbers. In addition, $\alpha$ and  $\beta$ are set to be not less than 1 (i.e., $\alpha, \beta \geq 1$) in the work, which are modeled by $\emph{softplus}$ and then add a constant 1 \cite{chou2017improving}.

\subsubsection{Algorithm Workflow}
\par \figref{fig:framework_of_HATRPO_UCB} displays the framework of the proposed HATRPO-UCB algorithm, which has two phases that are the training phase and the implementation phase. Each UAV as an agent has a replay buffer and two networks (i.e., the actor network and critic network). The replay buffer is used to collect the data about the interaction between the UAV and the environment. Then, a central server is deployed with the replay buffers and networks of all agents for training. 

\begin{algorithm}[t]
	\caption{Heterogeneous-Agent Trust Region Policy Optimization for UCBMOP (HATRPO-UCB)}
	\label{alg:1}
	\hspace*{0.02in} {\bf Input:} 
	Number of episodes $N^{\mathrm{Epi}}$, number of agents $N$, \\
	\hspace*{0.02in} batch size $B$, stepsize $\alpha$,
	possible steps in line search $L$, line  \\
	\hspace*{0.02in} search acceptance threshold $\kappa$. \\
	\hspace*{0.02in} {\bf Initialize:} 
	Actor networks $\{\theta^{0}_{i}, \forall i\in \mathcal{N}\}$, Critic networks \\ \hspace*{0.02in} $\{\phi^{0}_{i}, \forall i\in \mathcal{N}\}$, Replay buffers $\{\mathcal{B}_{i}, \forall i\in \mathcal{N}\}$. 
	\begin{algorithmic}[1]
		\For{episode = $1,\dots,N^{\mathrm{Epi}}$} 
		\State Initialize the environment, acquire local observations 
		\Statex \hspace*{0.16in} and global states of all agents $\{(o_{i}, s_{i}), \forall i\in \mathcal{N}\}$.
		\State Each UAV selects its own action $a_{i} = \pi_{\theta^{k}_{i}}(o_{i})$ from 
		\Statex \hspace*{0.16in} the Beta distribution.
		\State All UAVs execute actions to complete CB and then 
		\Statex \hspace*{0.16in} receive their rewards $\{r_{i}, \forall i\in \mathcal{N}\}$.
		\State Push transition of each agent $\langle  o_{i}, s_{i}, a_{i}, r_{i}\rangle$ into \Statex \hspace*{0.16in} its own replay buffer $\mathcal{B}_{i}$.
		\If{the training condition is met}
		\State Draw a random permutation of agents $i_{1:N}$.
		\For{agent $i=1,\dots, N$}
		\State Sample a random mini-batch of $B$ transitions 
		\Statex \hspace*{0.57in} from $\mathcal{B}_{i}$.
		\State Compute advantage function $A_{i}(s_{i}, \boldsymbol{a})$ based 
		\Statex \hspace*{0.57in} on the critic network.
		\State Obtain the compound policy ratio of the 
		\Statex \hspace*{0.57in} previous agents  $\prod^{i-1}_{l=1}\frac{\pi_{\theta^{k+1}_{l}}(a_{l}|o_{l})}{\pi_{\theta^{k}_{l}}(a_{l}|o_{l})}$, set 
		\Statex \hspace*{0.57in} $M_{i}(s_{i}, \boldsymbol{a})=A_{i}(s_{i}, \boldsymbol{a})\prod^{i-1}_{l=1}\frac{\pi_{\theta^{k+1}_{l}}(a_{l}|o_{l})}{\pi_{\theta^{k}_{l}}(a_{l}|o_{l})}$.
		\State Use $M_{i}(s_{i}, \boldsymbol{a})$ to estimate the gradient of 
		\Statex \hspace*{0.57in} maximization objective of agent $\boldsymbol{g}^{k}_{i}$ according 
		\Statex \hspace*{0.57in} to \eqaref{eq:the_computation_of_gradient}.
		\State Compute the Hessian of the expected 
		\Statex \hspace*{0.57in} KL-divergence $\boldsymbol{H}^{k}_{i}$ to approximate the KL 
		\Statex \hspace*{0.57in} constraint.
		\State $\boldsymbol{g}^{k}_{i}$ and $\boldsymbol{H}^{k}_{i}$ are introduced to update the 
		\Statex \hspace*{0.57in} parameter of actor network $\theta^{k+1}_{i}$ according to \Statex \hspace*{0.57in} \eqaref{eq:closed_form_policy_update_of_HATRPO}.
		\State Update critic network by the following 
		\Statex \hspace*{0.57in} equation:
		\Statex \hspace*{0.57in} $\phi^{k+1}_{i} = \operatorname{argmin}_{\phi_{i}}\frac{1}{B}\sum\limits^{B}_{b=1} (V_{\phi_{i}}(s^{b}_{i})- R^{b}_{i})^2$,
		\Statex \hspace*{0.57in} where $R^{b}_{i}$ is the discounted return at state $s^{b}_{i}$.
		\EndFor
		\EndIf
		\EndFor
	\end{algorithmic}
	\hspace*{0.02in} {\bf Output:} Trained models of all agents.
\end{algorithm}

\par In the beginning, the central server will initialize the parameters of networks. Then, at each time slot, the central server will exchange the information with UAVs and then control them, where the amount of the information is very small because it only includes the observation information of UAVs and actions required to be done by UAVs, and the communication distance is very short compared with that between the UAVs and the BS. Hence, for performing the communication tasks, the above communication overhead is very small and acceptable. Then, for illustrating the specific process at each time slot, taking the $i$-th agent as an example, after the central server receives the information, the actor network $\pi_{\theta_{i}}$ utilizes the observation $o_i$ to generate a Beta distribution about action, and then an action $a_i$ is sampled from the Beta distribution and sent to the UAV. The critic network is used to predict the Q-value of current global state $s_i$. After executing the sampled action, the $i$-th agent receives a reward $r_i$ from the environment and observes the next local state $o^{\prime}_{i}$ and global state $s^{\prime}_{i}$. After that, the above information tuple $\langle o_i, s_i, a_i, r_i\rangle$ will be stored in the replay buffer $\mathcal{B}_i$. 

\par All agents will execute more time slots until their replay buffers accumulate enough data, and then the replay buffers will randomly select a mini-batch of tuples for training networks. The sequential updating scheme is adopted to update policies of all agents. In each training, all agents are updated sequentially in a randomly generated order. For the $i$-th agent, the first step is to estimate the gradient $\boldsymbol{g}^{k}_{i}$, before that, the advantage function of $i$-th agent $A_{i}(s_{i}, \boldsymbol{a})$ and the compound policy ratio of the previous agents $\prod^{i-1}_{l=1}\frac{\pi_{\theta^{k+1}_{l}}(a_{l}|o_{l})}{\pi_{\theta^{k}_{l}}(a_{l}|o_{l})}$ are required to compute. Then, the Hessian of the expected KL-divergence $\boldsymbol{H}^{k}_{i}$ is computed to approximate the KL constraint. After completing the computation of $\boldsymbol{g}^{k}_{i}$ and $\boldsymbol{H}^{k}_{i}$, the parameter of actor network $\theta_{i}$ can be updated via backtracking line search. As for the critic network $V_{\phi_{i}}$, the update of network parameter can be achieved by optimizing the following loss function: $\frac{1}{B}\sum^{B}_{b=1}(V_{\phi_{i}}(s^{b}_{i})-R^{b}_{i})^{2}$. Therefore, the above updating procedure will be repeated until HATRPO-UCB converges. The whole training process can be seen in \textbf{Algorithm \ref{alg:1}}.

\par During the implementation stage, only the trained actor network is deployed to the UAV. In each communication mission, all UAVs receive the local observations from the environment, and then select actions through the actor networks. Afterwards, according to the selected actions, all UAVs move to the target locations and adjust the excitation current weights for performing CB. Compared with the training stage, the implementation stage is completed online and does not need much computational resource. In contrast, the training stage costs much computational resource, which is conducted offline in a central server.

\subsubsection{Analysis of HATRPO-UCB}
\par In this section, the computational complexity of the training and implementation stage of the proposed HATRPO-UCB is discussed. 

\par \textbf{Complexity of training}. For $n_{\mathrm{uav}}$ agents, each agent is equipped with an actor network and a critic network that are formed by DNNs. A DNN consists of an input layer, $L$ fully connected layers and an output layer, where $z_{0}$, $z_{i}$ and $z_{L+1}$ denote the number of neurons of input layer, $i$-th fully connected layer and output layer, respectively. Then, the computational complexity of DNN at each time slot is $O(\sum^{L+1}_{i=1}z_{i-1}\cdot z_{i})$. 

\par At each training process, a mini-batch of data with $n_{\mathrm{epi}}$ episodes is sampled from the replay buffer, but each episode only contains one time slot. Additionally, the critic network is updated by the stochastic gradient methods, thus the total computational complexity of critic network is $O(n_{\mathrm{epi}}\cdot(\sum^{L+1}_{i=1}z_{i-1}\cdot z_{i}))$. However, the actor network is updated using the backtracking line search. In \eqaref{eq:closed_form_policy_update_of_HATRPO}, the gradient $\boldsymbol{g}^{k}_{i}$ is first computed, whose computational complexity is $O(n_{\mathrm{epi}}\cdot (\sum^{L+1}_{i=1}z_{i-1}\cdot z_{i}))$. Then the Hessian of expected KL-divergence $\boldsymbol{H}^{k}_{i}$ is computed to approximate the KL constraint. The computational complexity of expected KL-divergence is $O(n_{\mathrm{epi}}\cdot (\sum^{L+1}_{i=1}z_{i-1}\cdot z_{i}))$, and the computational one of Hessian matrix is $O((\sum^{L+1}_{i=1}z_{i-1}\cdot z_{i})^2)$. Hence, the computational complexity of  $\boldsymbol{H}^{k}_{i}$ is $O(n_{\mathrm{epi}}\cdot (\sum^{L+1}_{i=1}z_{i-1}\cdot z_{i}) + (\sum^{L+1}_{i=1}z_{i-1}\cdot z_{i})^2)$. For computing $(\boldsymbol{H}^{k}_{i})^{-1}\boldsymbol{g}^{k}_{i}$, the conjugate gradient algorithm is adopted, which obtains result by searching iteratively until convergence. The computational complexity of each iteration is $O((\sum^{L+1}_{i=1}z_{i-1}\cdot z_{i})^2)$. Then the total computational complexity of $(\boldsymbol{H}^{k}_{i})^{-1}\boldsymbol{g}^{k}_{i}$ is $O(n_{\mathrm{epi}}\cdot (\sum^{L+1}_{i=1}z_{i-1}\cdot z_{i}) + l_{\mathrm{conj}}\cdot (\sum^{L+1}_{i=1}z_{i-1}\cdot z_{i})^2)$, where $l_{\mathrm{conj}}$ is the number of iterations. During the backtracking line search, the computational complexity is $O(l_{\mathrm{bt}}\cdot n_{\mathrm{epi}}\cdot (\sum^{L+1}_{i=1}z_{i-1}\cdot z_{i}))$, where $l_{\mathrm{bt}}$ is the search number and $O(n_{\mathrm{epi}}\cdot (\sum^{L+1}_{i=1}z_{i-1}\cdot z_{i}))$ is the computational complexity in each search. Hence, the total computational complexity of actor network is $O(l_{\mathrm{bt}}\cdot n_{\mathrm{epi}}\cdot(\sum^{L+1}_{i=1}z_{i-1}\cdot z_{i})+l_{\mathrm{conj}}\cdot (\sum^{L+1}_{i=1}z_{i-1}\cdot z_{i})^2)$. Assume that HATRPO-UCB needs to be trained $l_{\mathrm{t}}$ times for convergence, the overall computational complexity in the training phase is $O(l_{\mathrm{t}}\cdot n_{\mathrm{uav}}\cdot (l_{\mathrm{bt}}\cdot n_{\mathrm{epi}}\cdot (\sum^{L+1}_{i=1}z_{i-1}\cdot z_{i})+l_{\mathrm{conj}}\cdot (\sum^{L+1}_{i=1}z_{i-1}\cdot z_{i})^2))$.

\par \textbf{Complexity of inference}. In the implementation phase, all UAVs only use the trained actor networks to make decisions. Therefore, the computational complexity of HATRPO-UCB in the implementation stage is $O(n_{\mathrm{uav}}\cdot (\sum^{L+1}_{i=1}z_{i-1}\cdot z_{i}))$.

\begin{remark}
Note that determining the theoretical bounds and convergence of the proposed MADRL-based algorithm confronts significant challenges~\cite{Zhou2023}. Specifically, MADRL involves tuning numerous hyperparameters, such as learning rates, network architectures, and exploration strategies, resulting in an explosion of potential combinations. Thus, analyzing theoretical bounds and convergence becomes nearly infeasible given the vast parameter space. Moreover, DNNs introduce approximation errors when modeling complex functions, particularly in high-dimensional state spaces of the formulated optimization problem. These errors can lead to suboptimal performance in certain states, complicating the analysis of theoretical bounds and convergence. In addition, MADRL models interact with environments that introduce uncertainty factors, including noise and randomness. These uncertainties can yield different outcomes in various interaction trajectories, further complicating the analysis of the theoretical bounds and convergence~\cite{Zhang2021}. Thus, we evaluate the performance of the proposed HATRPO-UCB by conducting extensive simulations like the works in~\cite{Li2022a, Feriani2021} in the following. 
\end{remark}

\section{Simulation Results}
\label{sec:simulation_results_and_analysis}
In this section, we evaluate the performance of the proposed HATRPO-UCB algorithm for UCBMOP. 

\subsection{Simulation Configuration}
\par We implement simulations in an environment with Python 3.8 and Pytorch 1.10, and perform all experiments on a server with AMD EPYC 7642 48-Core CPU, NVIDIA GeForce RTX 3090 GPU and 128GB RAM. 

\par In the simulation, a 100 m $\times$ 100 m square area is considered, where 16 UAVs are flying in the area. The mass ($m_{UAV}$), the minimum and maximum flying heights ($H_{min}$ and $H_{max}$) of each UAV are set as 2 kg, 100 m and 120 m~\cite{sun2021time}, respectively. The minimum distance between any UAVs ($d_{min}$) is 0.5 m. Moreover, the transmit power of each UAV, the carrier frequency ($f_c$) and the total noisy power spectral density are 0.1 W, 2.4 GHz and -157 dBm/Hz, respectively \cite{mozaffari2018communications} \cite{li2021physical}. The path loss exponent ($\alpha$), as well as the attenuation factors of LoS and NLoS links ($\mu_{LoS}$, $\mu_{NLoS}$) are 2, 3 dB and 23 dB, respectively \cite{mozaffari2017wireless}. Other parameter configurations about system model are summarized in \tabref{table:parameter_setting}. Based on these settings and the mathematical models shown in Section~\ref{sec:system_model_and_problem_formulation}, we build a simulation environment for the DRL model to interact and collect data.

\par In the MADRL stage, an environment is built according to the UCBMOP. The MADRL algorithm learns by interacting with the environment. During the training stage, the environment will run $4\times10^5$ episodes, where each episode only has one time slot. For the settings of reward weight coefficients, we first set the weight coefficients by normalizing each reward to make the rewards on the same order of magnitude. Then, we fine-tune the reward weights by considering their importance and characteristics (the tuning process and results are shown in Appendix A of the supplemental material). As such, the reward weight coefficients $\omega_1$, $\omega_2$, $\omega_3$, $\omega_4$, and $\omega_5$ are set to 100, 4, 30, 12 and 5, respectively.

\par In HATRPO-UCB, all actor and critic networks are three-layer fully connected neural networks (i.e., two hidden layers and one output layer). Each hidden layer contains 64 neurons, which is initialized by orthogonal initialization, and it is equipped with ReLU activation function. The actor network is updated by the backtracking line search, meanwhile, the KL-divergence between the new actor network and the old one must meet the KL-threshold which is set as 0.001. The Adam optimizer \cite{kingma2014adam} is employed to update the critic network, where the learning rate is 0.005. The value setting of other related parameters is also presented in \tabref{table:parameter_setting}.

\begin{table}[t]
	\renewcommand\arraystretch{1.05}	
	\caption{Parameter settings}
	\centering
	\begin{tabular*}{\linewidth}{m{6.3cm}<{\raggedright} m{1.7cm}<{\raggedright}}
		\hline
		Parameter & Value \\  
		\hline
		Fuselage drag ratio ($d_0$) & 0.6\\
		Air density ($\rho$) & 1.225 $\mathrm{km/m^3}$ \\
		Rotor solidity ($s$) & 0.05  \\
		Rotor disc area ($A$) & 0.503 $\mathrm{m^2}$ \\
		Tip speed of the rotor blade ($v_{tip}$) & 120 $\mathrm{m/s}$\\
		Mean rotor induced velocity in hover ($v_0$) & 4.03  \\
		Blade profile power ($P_B$) & 79.76 \\
		Induced power ($P_I$) & 88.66  \\
		Discount factor ($\gamma$) & 0.99 \\
		Number of line searches & 10 \\
		Clip parameter for loss value & 0.2 \\
		Max norm of gradients & 10 \\
		Accept ratio of loss improve & 0.5\\
		Weight initialization gain for actor network  & 0.01 \\
		Weight initialization for neural network  & Orthogonal \\
		\hline
	\end{tabular*}
	
	\label{table:parameter_setting}
\end{table}

\par In addition, to verify the effectiveness and performance of HATRPO-UCB, except for conventional HATRPO, the following baseline methods are also introduced for comparison. 
\begin{itemize}
	\item \textbf{LAA}: All UAVs form a linear antenna array (LAA), and they are symmetrically excited and located about the origin of the array \cite{mozaffari2018communications}.
	\item \textbf{RAA}: A rectangular antenna array (RAA) is formed by all UAVs, where all UAVs are also symmetrically excited and located about the origin of the array.
	\item \textbf{MADDPG}: Multi-Agent Deep Deterministic Policy Gradient (MADDPG) is based on the centralized training with decentralized execution (CTDE) paradigm, extending Deep Deterministic Policy Gradient (DDPG) to MADRL \cite{lowe2017multi}. 
	\item \textbf{IPPO}: Independent Proximal Policy Optimization (IPPO) is that the single-agent PPO algorithm is adopted directly to solve multi-agent tasks \cite{de2020independent}. The IPPO has been justified that it can achieve excellent performance on SMAC.
	\item \textbf{MAPPO}: Multi-Agent Proximal Policy Optimization (MAPPO) is an extension of PPO \cite{yu2021surprising}. MAPPO utilizes parameter-sharing trick, then all agents jointly use a policy network and a value network. The global state is the input of the value network.
\end{itemize}

\par Note that the proposed $\emph{observation enhancement}$ technique are added to these abovementioned MADRL algorithms, so that making sure that they can solve the UCBMOP properly. Furthermore, they are also tuned continuously and achieve the convergence.

\subsection{Convergence Analysis}
\par
\begin{figure}[t]
	\centering
	\includegraphics[width=7cm]{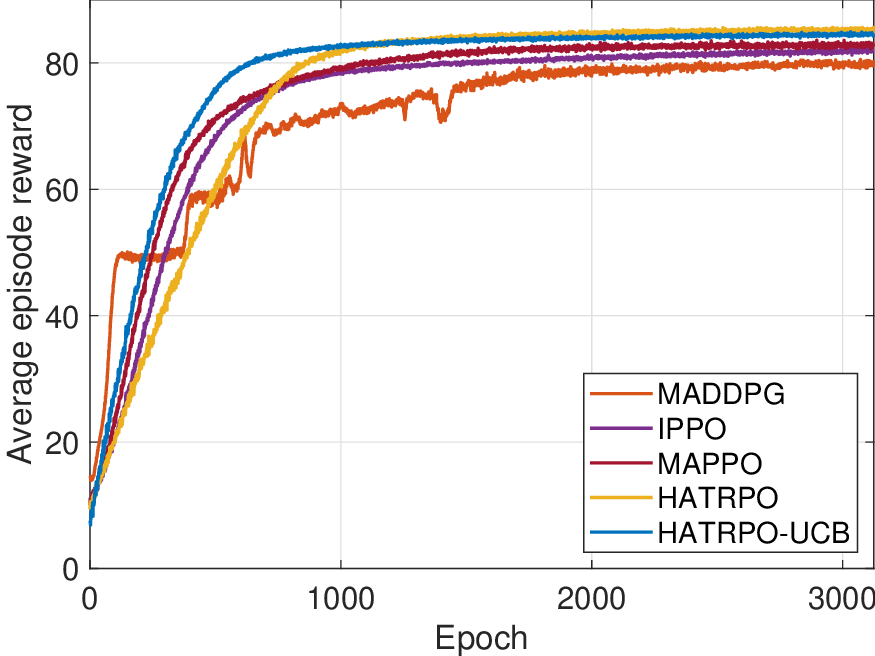}
	\caption{Convergence performance of different methods.}
	\label{fig:convergence_performance}
\end{figure}

 In practice, DRL-based models are usually deployed only after achieving convergence through training. Even if the practical environment changes dynamic, we also need to re-train and fine-tune the model before deploying it. Thus, the convergence performance of HATRPO-UCB is vital. As such, in the section, we compare the convergence performance of HATRPO-UCB with other algorithms. \figref{fig:convergence_performance} shows the convergence performance of all algorithms. It can be observed that all algorithms converge successfully. HATRPO-UCB achieves the fastest convergence at approximately 750 epoch, then the second algorithm is HATRPO. MAPPO, IPPO and MADDPG are slowest, which all begin to converge at about 1400 epoch. However, in the performance, the reward obtained by HATRPO is higher than that of the HATRPO-UCB, reaching about 85. IPPO and MAPPO are both worse than HATRPO-UCB. MADDPG has the poorest performance, which only converges to around 80. The reasons are that the introduced observation enhancement provides representative and adaptive information for the agents to take action, the agent-specific global state is able to improve the fitting performance of critic learning, and the Beta distribution overcomes the boundary effect issue. Such improvements could effectively address the major issues in the training process, thus facilitating the algorithm in achieving reliable and high-performance outcomes and converging rapidly.

\begin{figure}[t]
	\centering
	\hspace{-0.4cm}
	\includegraphics[width=7cm]{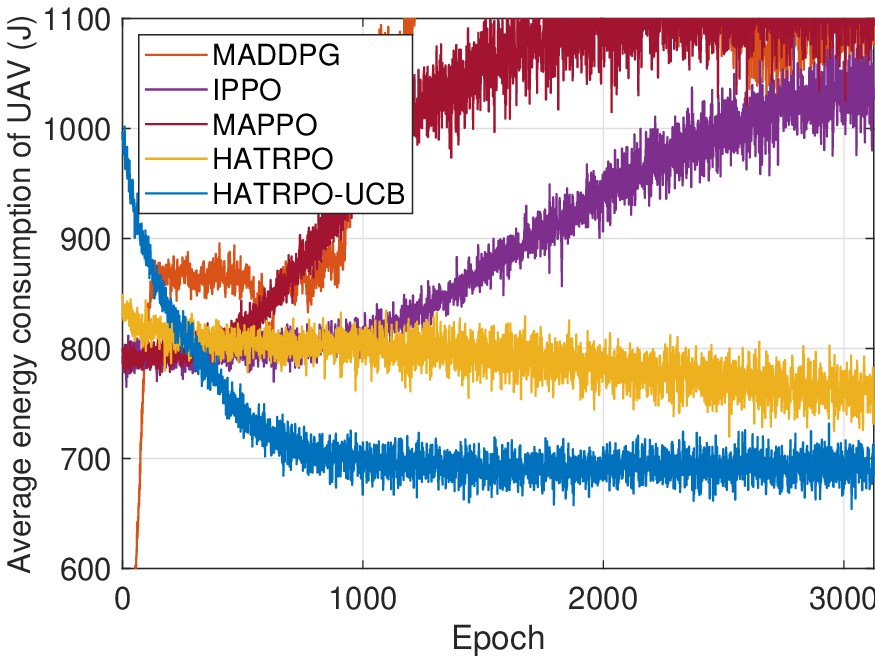}
	\caption{Optimization results of the UAV energy consumption obtained by different methods.}
	\label{fig:optimization_of_energy_consumption}
\end{figure}

\begin{table*}[htbp]
	\aboverulesep=0pt
	\belowrulesep=0pt
	\renewcommand\arraystretch{1.2}
	\caption{Numerical optimization results of different methods}
	\centering
	\begin{tabular}{m{2.1cm}<{\centering}m{2.4cm}<{\centering} m{2.4cm}<{\centering} m{2.4cm}<{\centering}  m{2.4cm}<{\centering}}
		\hline
		& \multicolumn{2}{c}{First BS} & \multicolumn{2}{c}{Second BS} \\ \cmidrule(lr){2-3} \cmidrule(lr){4-5}
		Method&Transmission rate (\textbf{bps})&Energy consumption (\textbf{J})&Transmission rate (\textbf{bps})&Energy consumption (\textbf{J}) \\
		\hline
		LAA & $1.020\times10^6$ & 15208 & $1.793\times10^7$  & / \\
		RAA & $0.998\times10^6$ & 14217 &  $1.788\times10^7$ & / \\
		MADDPG & $1.027\times10^6$ & 22319 & $\boldsymbol{1.904\times10^7}$ & 17913  \\
		IPPO & $1.022\times10^6$ & 15790 & $1.867\times10^7$ & 21110 \\
		MAPPO & $\boldsymbol{1.035\times10^6}$ & 15025 & $1.896\times10^7$  & 23252 \\
		HATRPO & $1.032\times10^6$ & 13765 & $1.838\times10^7$ & 11779 \\
		HATRPO-UCB & $1.029\times10^6$ & $\boldsymbol{13401}$ & $1.825\times10^7$  & $\boldsymbol{10261}$ \\
		\hline
	\end{tabular}
	
	\label{table:optimization_results}
\end{table*}

\begin{figure*}[t]
	\centering
	\hspace{-0.2cm}
	\subfloat[]{
		\centering
		\includegraphics[width=5.4cm]{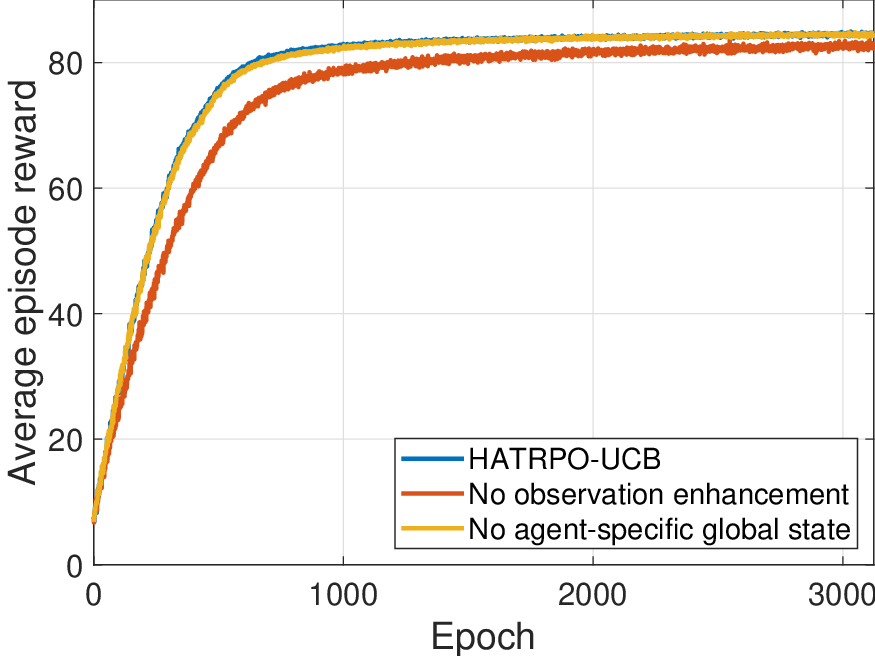}
		\label{fig:ablation_analysis_a}
	}
	\hspace{0.1cm}
	\subfloat[]{
		
		\centering
		\includegraphics[width=5.6cm]{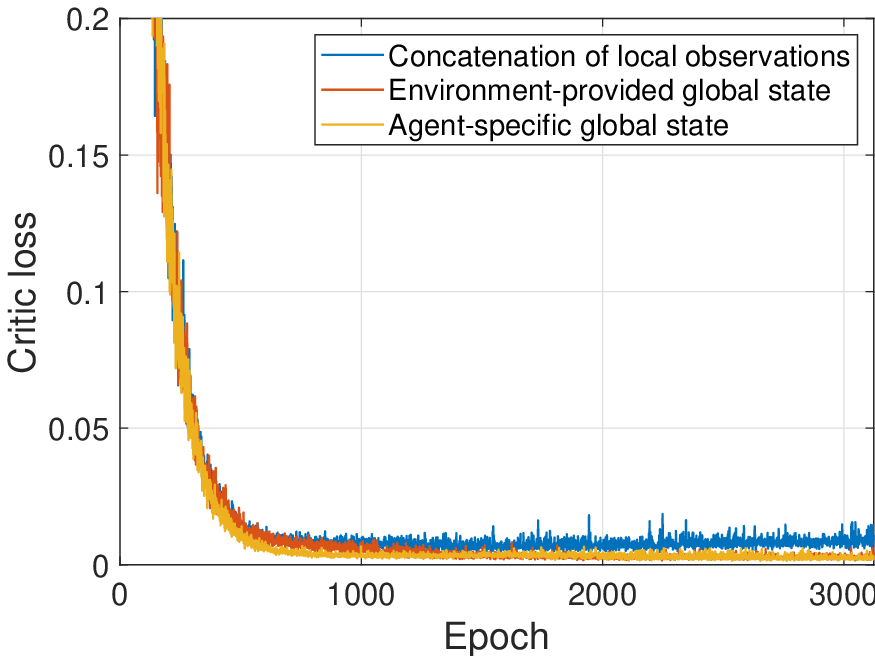}
		\label{fig:ablation_analysis_b}
	}
	\hspace{0.1cm}
	\subfloat[]{
		\centering
		\includegraphics[width=5.4cm]{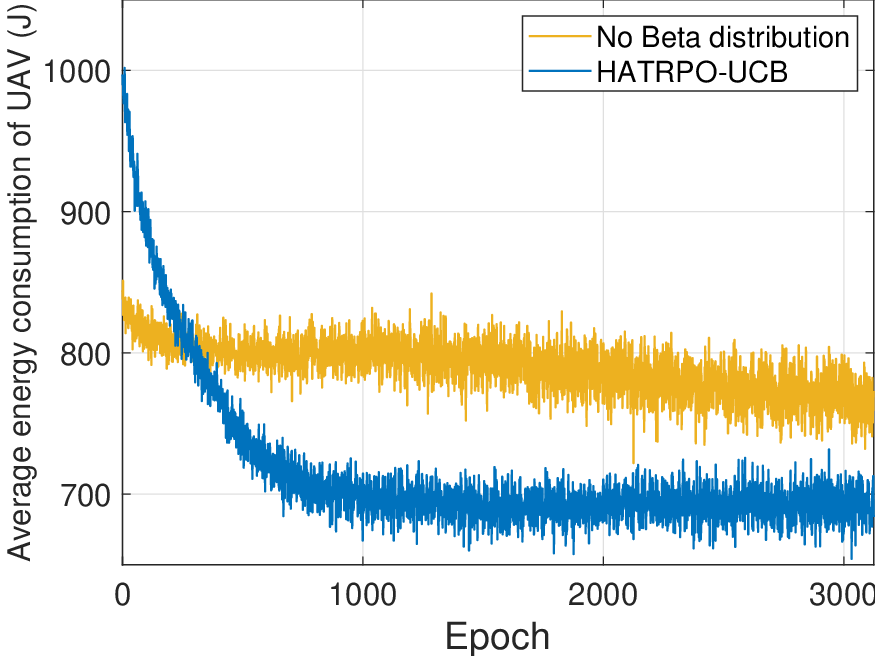}
		\label{fig:ablation_analysis_c}
	}
	\caption{Effectiveness of different techniques. ($\emph{Observation enhancement}$, $\emph{agent-specific global state}$ and $\emph{Beta distribution for policy}$).}
	\label{fig:ablation_analysis}
\end{figure*}

\par To further analyze these algorithms, we also depict the optimization process of all alogrithms to the UAV energy consumption in \figref{fig:optimization_of_energy_consumption}. It can be observed that HATRPO-UCB obtains the best optimization performance, making the energy consumption of UAV reduce to about 700 J and achieving the convergence at approximately 1000 epochs. Then, HATRPO has the certain optimization performance, but it does not have the obvious optimization effect like HATRPO-UCB and the ultimate optimization result is not better than that of HATRPO-UCB. In addition, MAPPO, IPPO and MADDPG have no optimization effect, whose optimization performance becomes worse continously with the training process. The more details about performance of all algorithms will be introduced in the following sections.

\subsection{Performance Comparison}
\par The practical performance of all approaches is compared in the section. In our system, all UAVs have other kinds of tasks after completing CB, but they may continue to perform CB for communicating with another BS. Thus, we will analyze the actual performance of UVAA communicating continuously with two BSs. \tabref{table:optimization_results} shows the numerical optimization results of various methods. It can be observed that HATRPO-UCB achieves the best performance on the optimization of energy consumption of UVAA, and obtains the outstanding results on the transmission rate optimization. The highest transmission rates to the two BSs are achieved by MAPPO and MADDPG, respectively, but they consume more energy. Furthermore, remaining MADRL approaches also outperform HATRPO-UCB in terms of the optimization of transmission rate. The reason may be that all UAVs optimized by these algorithms fly higher and closer to BS, thus the UVAA can get the lower path loss and shorter distance with BS. However, the more energy consumption will be induced.

\par We also show the flight paths of UAVs in the Appendix B of supplemental material. Benefiting from the DNN, the UVAA optimized by MADRL methods can directly perform CB for the second BS without consuming the much computation resource. Besides, we can also observe that all UAVs optimized by other MADRL approaches can achieve the higher altitude and the shorter distance with BS compared with HATRPO-UCB, and thus, they can obtain the higher transmission rate but will consume more energy as shown in \tabref{table:optimization_results}. In summary, HATRPO-UCB learns the best strategy, making UVAA obtain the great transmission rate and save the energy at the same time.

\subsection{Ablation Analysis}
\par
\begin{figure}[t]
	\centering
	\subfloat[]{
		\centering
		\includegraphics[width=4.2cm]{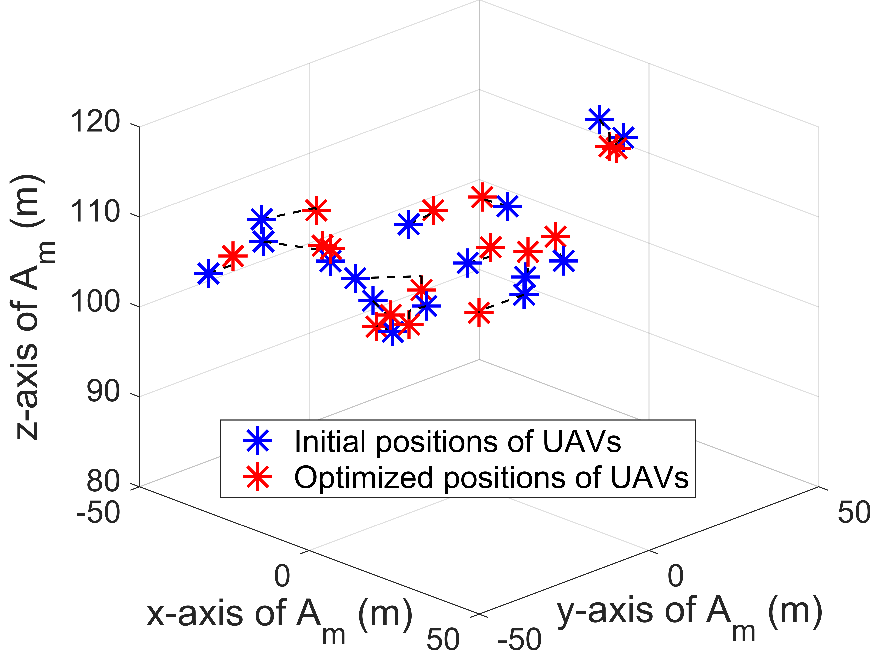}
		\label{fig:first_trick_a}
	}
	\subfloat[]{
		\centering
		\includegraphics[width=4.2cm]{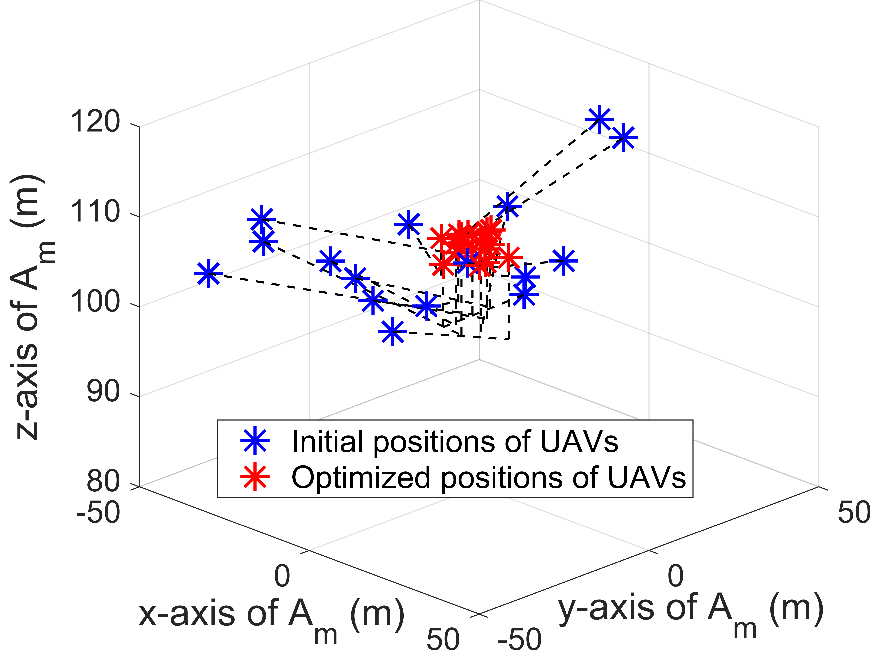}
		\label{fig:first_trick_b}
	}
	
	\caption{The impact of the first technique to UVAA when communicating with a BS. (a) Without \emph{observation} \emph{enhancement}; (b) With \emph{observation} \emph{enhancement}.}
	\label{fig:first_trick}
\end{figure}

 For verifying the effectiveness of proposed techniques, we conduct some ablation experiments.
\figref{fig:ablation_analysis_a} displays some convergence performance as HATRPO-UCB is not implemented with some techniques. When HATRPO-UCB is not equipped with \emph{observation enhancement}, we can observe that the performance is decreased. For further analysis, we present the practical performance of UVAA whether adopting \emph{observation enhancement} or not in \figref{fig:first_trick}. As can be seen, when using \emph{observation enhancement}, all UAVs can cooperate better, and they can also fly toward the BS to shorten the communication distance. On the contrary, without \emph{observation enhancement}, all UAVs are scattered and have no clear flight targets. Furthermore, the CB performance is also influenced. The UVAA gets 24.85 reward about the transmission rate that is smaller than 25.05 obtained by the algorithm with \emph{observation enhancement}. Therefore, the \emph{observation enhancement} technique can significantly enable UVAA to learn the better strategy.

\par For testing the effectiveness of \emph{agent-specific global state}, we replace the agent-specific global state, then use the environment-provided global state to evaluate the algorithm performance. As shown in \figref{fig:ablation_analysis_a}, HATRPO-UCB suffers from the performance degradation, which is because the information contained in the environment-provided global state is insufficient. Likewise, the critic learning is also influenced. \figref{fig:ablation_analysis_b} shows the learning curves of the critic network based on different global states, where the concatenation of all local observations is also added for comparison. It can be observed that the loss based on the agent-specific global state is lower than those under the other two global states. The reason is that the concatenation of local observations contains much redundant information, increasing the training burden, thus the learning process becomes very unstable and performance degrades. Then, the environment-provided global state also influences the training efficiency due to the insufficient information. However, according to the two figures, the influence produced by the two global states looks small. The reason could be that each episode only has one time slot in the UCBMOP, which distinctly differs from the regular MADRL tasks that require sequential decision-making. Hence, the role of critic network is not as significant as that in the regular tasks.

\par As for \emph{Beta distribution for policy}, an action sampled from the Beta distribution is not beyond its setting range, which is beneficial for the policy gradient estimation so that no bias occurs. 
We show the impact of \emph{Beta distribution for policy} to the optimization about the energy consumption of UAV in \figref{fig:ablation_analysis_c}. It can be seen that \emph{Beta distribution for policy} can make HATRPO-UCB optimize the UAV energy consumption well, reducing the energy consumption to around 700 J. However, without \emph{Beta distribution for policy}, the optimization effect is substantially weakened, and the ultimate performance is not better than the algorithm with \emph{Beta distribution for policy}. Therefore, according to \cite{chou2017improving}, adopting the Beta distribution can help the agent learn the better strategy compared with the Gaussian distribution.

\begin{figure}[t]
	\centering
	\hspace{-0.1cm}
	\subfloat[]{
		\centering
		\includegraphics[width=4.1cm]{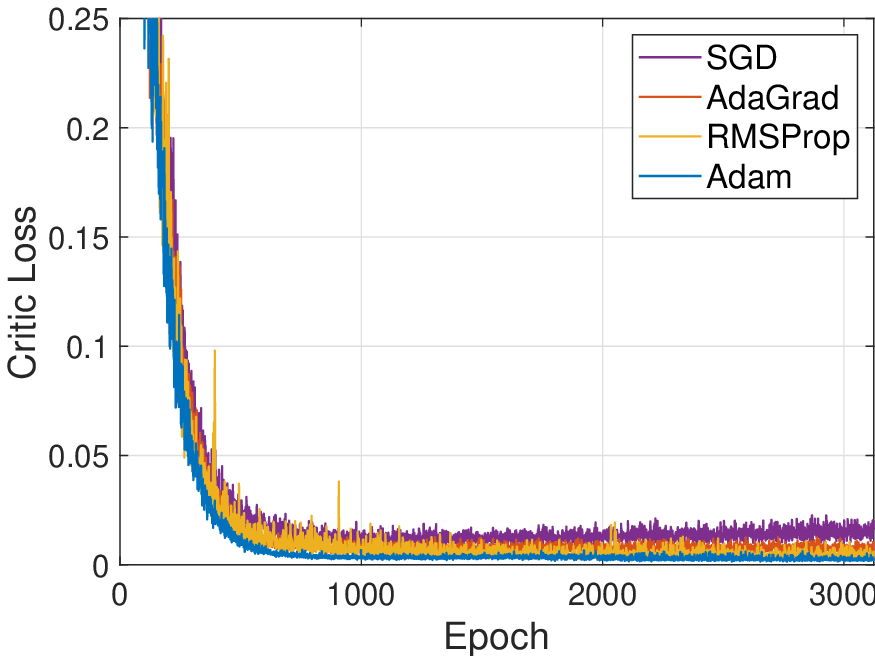}
		\label{fig:exp4_opt}
	}
	\subfloat[]{
		\centering
		\includegraphics[width=4.1cm]{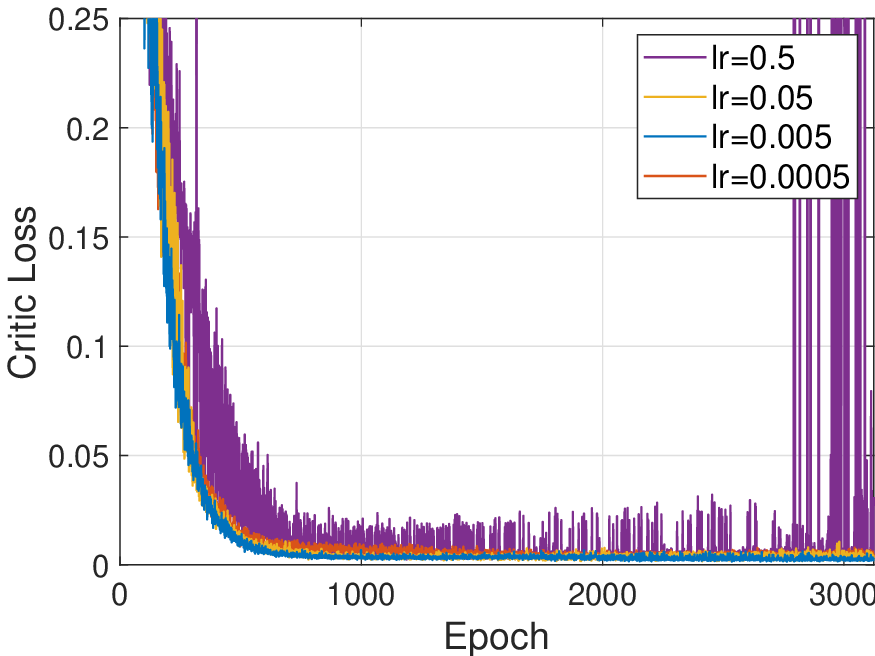}
		\label{fig:exp4_lr}
	}\\
	\hspace{-0.1cm}
	\subfloat[]{
		\centering
		\includegraphics[width=4.1cm]{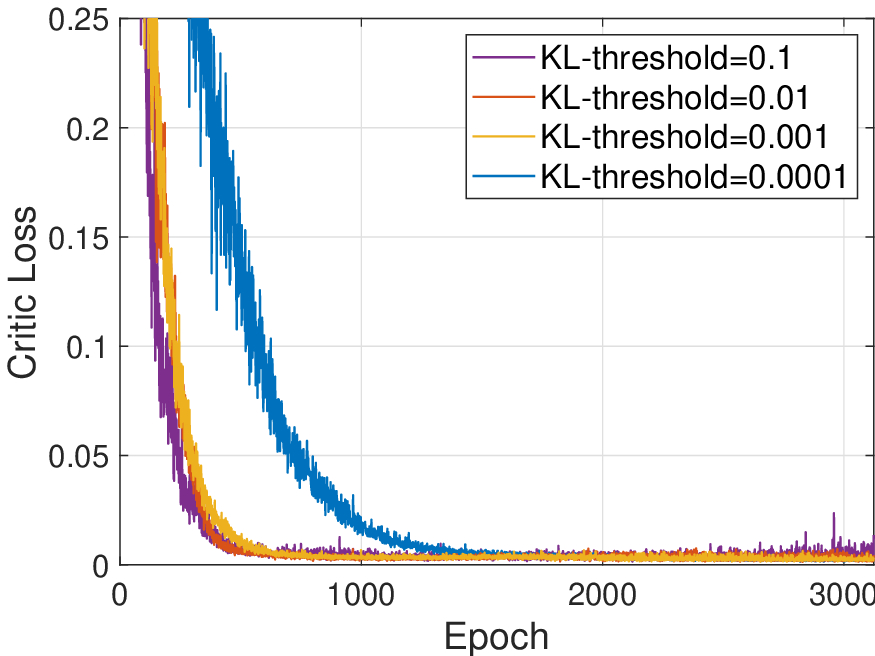}
		\label{fig:exp4_kl}
	}
	\subfloat[]{
		\centering
		\includegraphics[width=4.1cm]{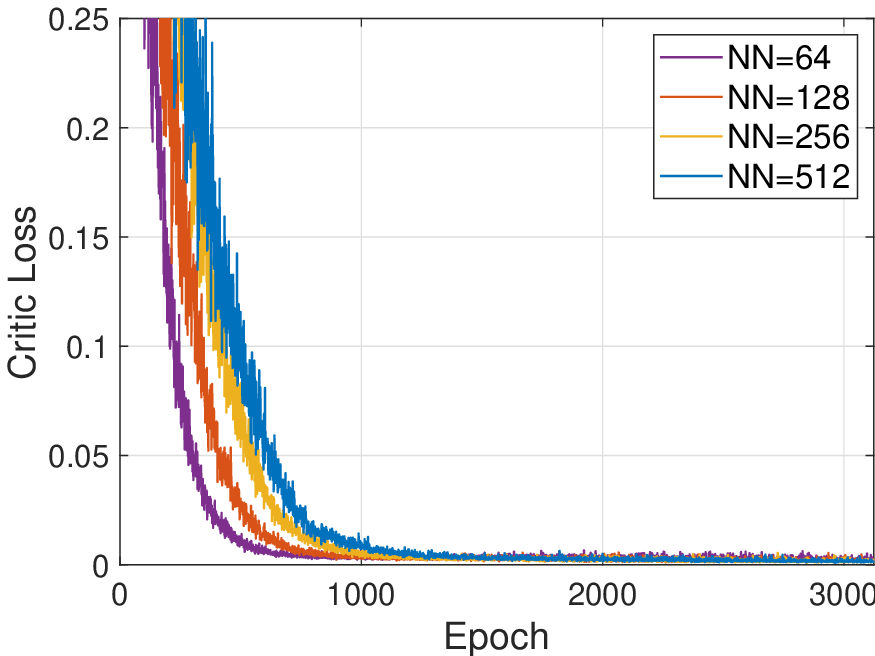}
		\label{fig:exp4_nn}
	}
	
	\caption{Impact of different hyperparameter settings.}
	\label{fig:hyperparameter_setting}
\end{figure}

\subsection{Impact of Hyperparameter Setting}
\par The appropriate hyperparameter setting is critical to the algorithm. In HATRPO-UCB, the following hyperparameters are closely related to the algorithm performance that are the Optimizer, Learning rate, KL-threshold and Number of neurons. Hence, we analyze the impact of different values of each hyperparameter to HATRPO-UCB and simultaneously verify the reasonableness of our setting. 

\par \textbf{Algorithm performance under different optimizers.} We first analyze the influence of various optimizers. Four optimizers are adopted to optimize the critic network, which are SGD\cite{bottou1991stochastic}, AdaGrad\cite{duchi2011adaptive}, RMSProp\cite{tieleman2012lecture} and Adam\cite{kingma2014adam}, respectively. Their optimization results about the critic loss are shown in \figref{fig:exp4_opt}. We can observe that, following the above order of optimizers, the optimized result becomes better. The performance of SGD is the worst, and the performance degradation appears in the following training. In the other optimizers, the training process also becomes unstable, resulting in an oscillating training curve. Then, when using Adam, the optimization performance is best and the training curve is smoothest. Therefore, Adam is choosen as the optimizer of HATRPO-UCB.

\par \textbf{Algorithm performance under different learning rates.} Except for determining the optimizer, the critic network is also optimized based on a certain learning rate. Hence, \figref{fig:exp4_lr} shows the learning process of critic network based on Adam under four learning rates. It can be observed that the optimization performance is best and the critic loss curve is smoothest as the learning rate is 0.005. When equaling to 0.05 or 0.0005, the curve oscillates slightly, which demonstrates that they are not appropriate under the current hyperparameter configuration. However, as the learning rate is 0.5, the training process gets very unstable and the performance degrades seriously at the end of training. The probable reason is that the learning rate is too large to produce the fluctuation.

\par \textbf{Algorithm performance under different KL-thresholds.} As for the KL-threshold, it is related to the update of actor network, but the KL-threshold also influences the learning speed of the critic network. \figref{fig:exp4_kl} describes the critic loss curve under various KL-thresholds.  We can observe that critic network begins to converge at approximately 1000 epoch as the KL-threshold is 0.001. Then, when the KL-threshold equals 0.1 and 0.01, the convergence speed of critic network becomes faster. However, these large KL-thresholds also cause the instability to the critic learning in the second half of the whole process, and then also make the actor learning unstable. At last, 0.0001 KL-threshold makes the critic network learn so slowly that it converges at about 2000 epoch. Therefore, 0.001 KL-threshold is the best choice, which can make the critic network converge properly.

\par \textbf{Algorithm performance under different neuron numbers.} The number of neurons of hidden layer has the important effect to the DNN. Hence, \figref{fig:exp4_nn} displays the performance of HATRPO-UCB under various numbers of neurons. We can observe that the convergence of critic network becomes slower as the number of neurons increases. The reason is that the large number of neurons makes the DNN big and complex, increasing the training burden. Hence, the number of neurons should be set to 64. In conclusion, according to all the above analysis, the hyperparameter setting with Adam, 0.005 learning rate, 0.001 KL-threshold and 64 neurons can make HATRPO-UCB achieve the best performance.

\begin{figure}[t]
	\centering
	\hspace{-0.2cm}
	\includegraphics[width=7.5cm]{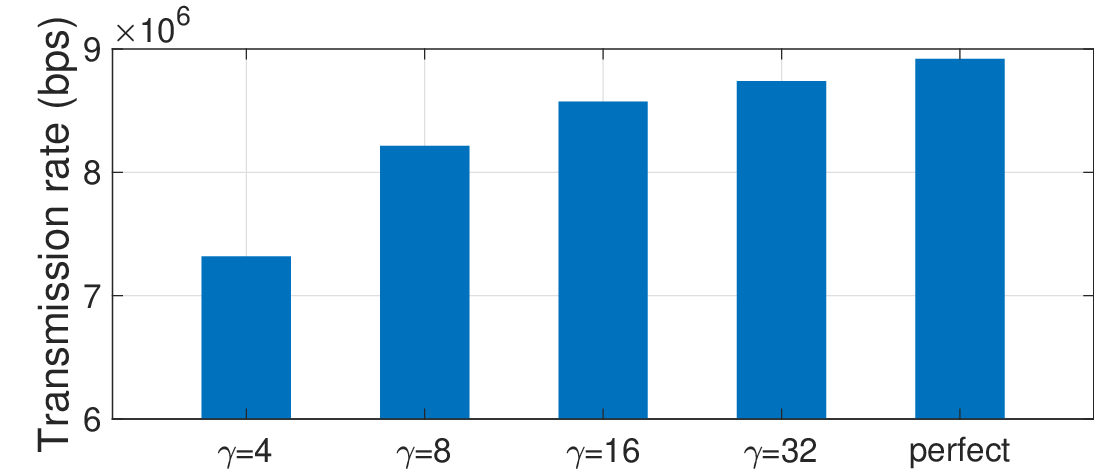}
	\caption{Transmission rate of UVAA under different phase errors.}
	\label{fig:phase_error}
\end{figure}

\subsection{Impact of Imperfect Synchronization}

\par The imperfect synchronization exists in CB, where the phase errors may be generated to influence the CB performance. Thus, we evaluate the impact of phase errors to the transmission performance of UVAA in the section. 

\par The phase error at $i$-th UAV antenna is denoted as $\epsilon_{i}$, and then the AF of UVAA can be rewritten as \cite{minturn2013distributed}
\begin{equation}\label{eq:array_factor_with_phase_error}
	\begin{aligned}
		AF(\theta, \phi) &=\\ &\sum_{i=1}^NI_{i}e^{j\Psi_{i}}e^{j[\frac{2\pi}{\lambda}(x_{i}\sin{\theta}\cos{\phi} + y_{i}\sin{\theta}\sin{\phi} + z_{i}\cos{\theta})+\epsilon_{i}]},
	\end{aligned}
\end{equation}
which $\epsilon_{i}$ is assumed to follow a Tikhonov (or von Mises) distribution \cite{minturn2013distributed} \cite{shmaliy2005mises}. The Tikhonov distribution is described as 
\begin{equation}\label{Tikhonov_distribution}
	\begin{aligned}
		f_{\epsilon}(\epsilon)=\frac{1}{2\pi I_{0}(\gamma)}e^{\gamma cos(\epsilon)},
	\end{aligned}
\end{equation}	
where $|\epsilon| < \pi$, $I_{0}(x)$ is the zero-th order modified Bessel function of the first kind, and $\gamma$ is the inverse of the variance of the phase error. Then, the impact of phase errors to the transmission rate of UVAA under different values of $\gamma$ is shown in \figref{fig:phase_error}. It can be seen that the phase errors make the UVAA transmission performance decline. However, the generated phase errors become smaller as $\gamma$ gets large, weakening the influence. For tackling the imperfect synchronization, there has been many works proposing various closed-loop or open-loop methods \cite{jayaprakasam2017distributed}. Furthermore, as the better synchronization algorithms are being proposed continuously, the impact of the imperfect synchronization will become less and less.

\section{Discussion}

\par In this section, we discuss possible issues and solutions when deploying the proposed method in actual scenarios.

\subsection{The Impact of UAV Collision}

\par In this part, we discuss the impact of UAV collisions. As shown in Eq.~\eqref{optimization_problem_6}, we set the minimum separation between two UAVs in UCBMOP. Following this, we set a penalty in the reward. Specifically, when any two UAVs violate the minimum separation constraint, they will get a tiny negative reward. As such, the proposed algorithm is able to make UAVs keep the minimum separation.

\par Moreover, many mature methods of UAV collision avoidance can also help the UAVs avoid violating the minimum separation in reality. For example, cameras, infrared, radar, LiDAR, sonar, etc, can be utilized by UAVs for obstacle detection. Based on this, the existing methods such as sense \& avoid~\cite{Zeng2018} can be appropriately embedded in the proposed optimization framework. This type of method focuses on reducing the computational cost with a short response time, deviating the UAVs from their original paths when needed, and then turning the UAVs to the previous flight state quickly. Since this method only consumes very little computational resources, energy, and time, it does not have a significant effect on the results of our work. As such, UAV collisions do not affect the effectiveness of the proposed method. 

\subsection{Additional Energy Consumption in Synchronization}

\par In the considered system, there is additional energy consumption in the synchronization process for beacon signal reception (for closed loop) and location estimation (for open loop). However, this additional energy consumption is minor and can be ignored, and the reasons are as follows.

\par \textbf{First}, the closed-loop synchronization methods are based on feedback, whose energy consumption is quite small compared to the transmission or motion energy consumption of the UAVs. Specifically, the closed-loop synchronization approaches mainly include two types that are the iterative bit feedback and rich feedback methods. In the considered UAV-enabled CB scenarios, both of these closed-loop methods result in negligible energy consumption. This is because the recently proposed methods, such as iterative bit feedback approaches and rich feedback methods, can achieve rapid convergence with minimal energy consumption (less than 1 Joule in most cases). As such, this additional energy consumption is much smaller than the motion energy of UAVs (often several hundred Joules each second).

\par
\begin{figure}
    \centering
        \subfloat[]{
       \includegraphics[width=0.49\linewidth]{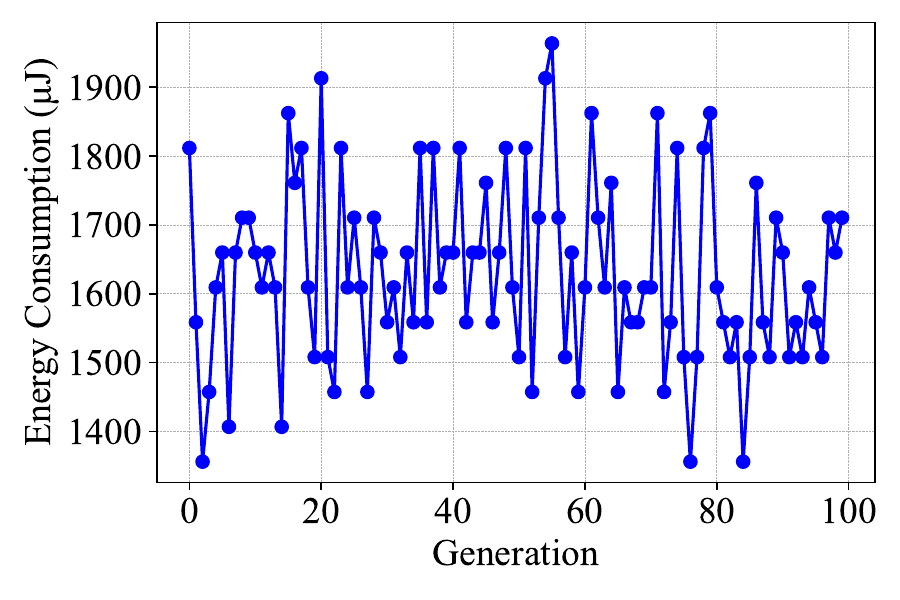}}
     \subfloat[]{
       \includegraphics[width=0.49\linewidth]{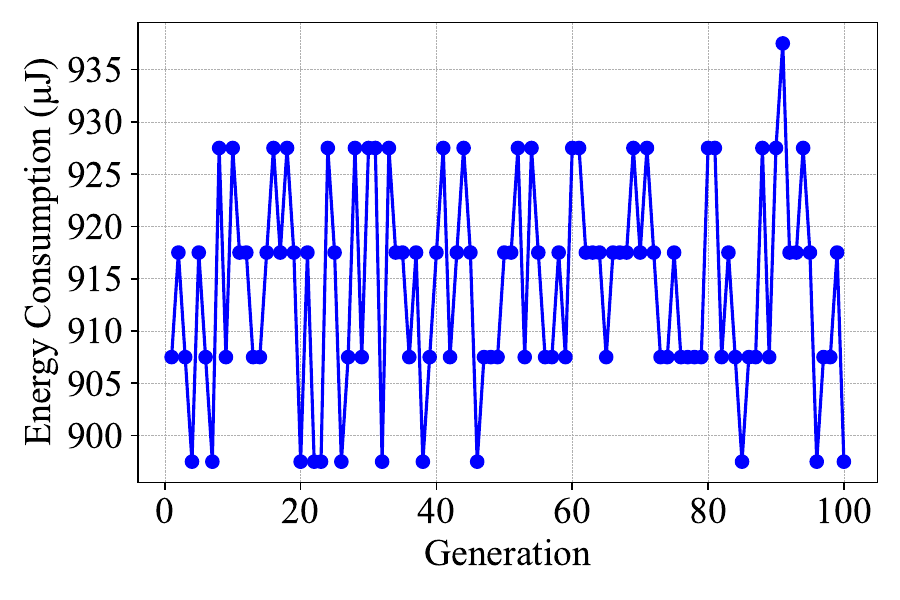}} \\
    \caption{Energy consumption of the UAVs for the synchronization process. (a) Closed loop. (b) Open loop. These synchronization methods need no more than one hundred times generation to complete synchronization tasks. As can be seen, the energy consumption of each generation is very small and can be ignored. }
    \label{fig:synchronization}
\end{figure}

 To demonstrate, we select a representative method, i.e., D1BF~\cite{Thibault2010}, to implement and evaluate. Following this, we introduce the energy consumption of point-to-point sending and broadcast receiving models from~\cite{Feng2009}. As shown in Fig.~\ref{fig:synchronization}(a), the energy consumption of the closed-loop synchronization is not larger than even 1 Joule. Thus, we can demonstrate that the additional energy consumption of closed-loop synchronization in the considered system is quite small and can be ignored.

\par \textbf{Second}, compared to the closed-loop synchronization methods, the open-loop synchronization methods are not based on iterative feedback and have less energy consumption. Specifically, the open-loop synchronization approaches can be divided into two categories that are the intra-node communication and blind method. On the one hand, the intra-node methods are mostly based on the master-slave architecture~\cite{Quitin2012}. We also use the energy model above and assume that the master node sends a reference beacon to all slave nodes and then the slaves synchronize their own phases. In this case, the energy consumption of synchronization of the UAVs can be shown in Fig.~\ref{fig:synchronization}(b). As can be seen, it is also not larger than 1 Joule and can be omitted. On the other hand, blind methods allow the nodes to synchronize among themselves and do not require feedback from the receiver or the reference nodes. Hence, the overhead of the blind method can be smaller. Thus, the additional energy consumption of open-loop synchronization in the considered system is also small and can be ignored.

\par Overall, the additional energy consumption for synchronization does not have an evident impact on the availability of the proposed method.

\section{Conclusion}
\label{sec:conclusion}
In this paper, a UAV-assistant A2G communication system is investigated, where multiple UAVs form a UVAA to perform CB for communicating with remote BSs. Then, we formulate a UCBMOP, aiming at simultaneously maximizing the transmission rate of UVAA and minimizing the energy consumption of all UAVs. Consider that the system is dynamic and the cooperation among UAVs is complex, we propose the HATRPO-UCB, which is an MADRL algorithm to address the problem. Except for combining conventional HATRPO, three techniques are proposed to enhance the performance of the proposed algorithm. Simulation results demonstrate that the proposed HATRPO-UCB learns the better strategy than other baseline methods including LAA, RAA, MADDPG, IPPO, MAPPO and conventional HATRPO, making UVAA achieve the best transmission performance and save the energy consumption simultaneously. Furthermore, the effectiveness of three techniques is also verified by the ablation experiments.

\bibliographystyle{ieeetr}
\bibliography{CB}

\begin{IEEEbiography}[{\includegraphics[width=1in,height=1.25in,clip,keepaspectratio]{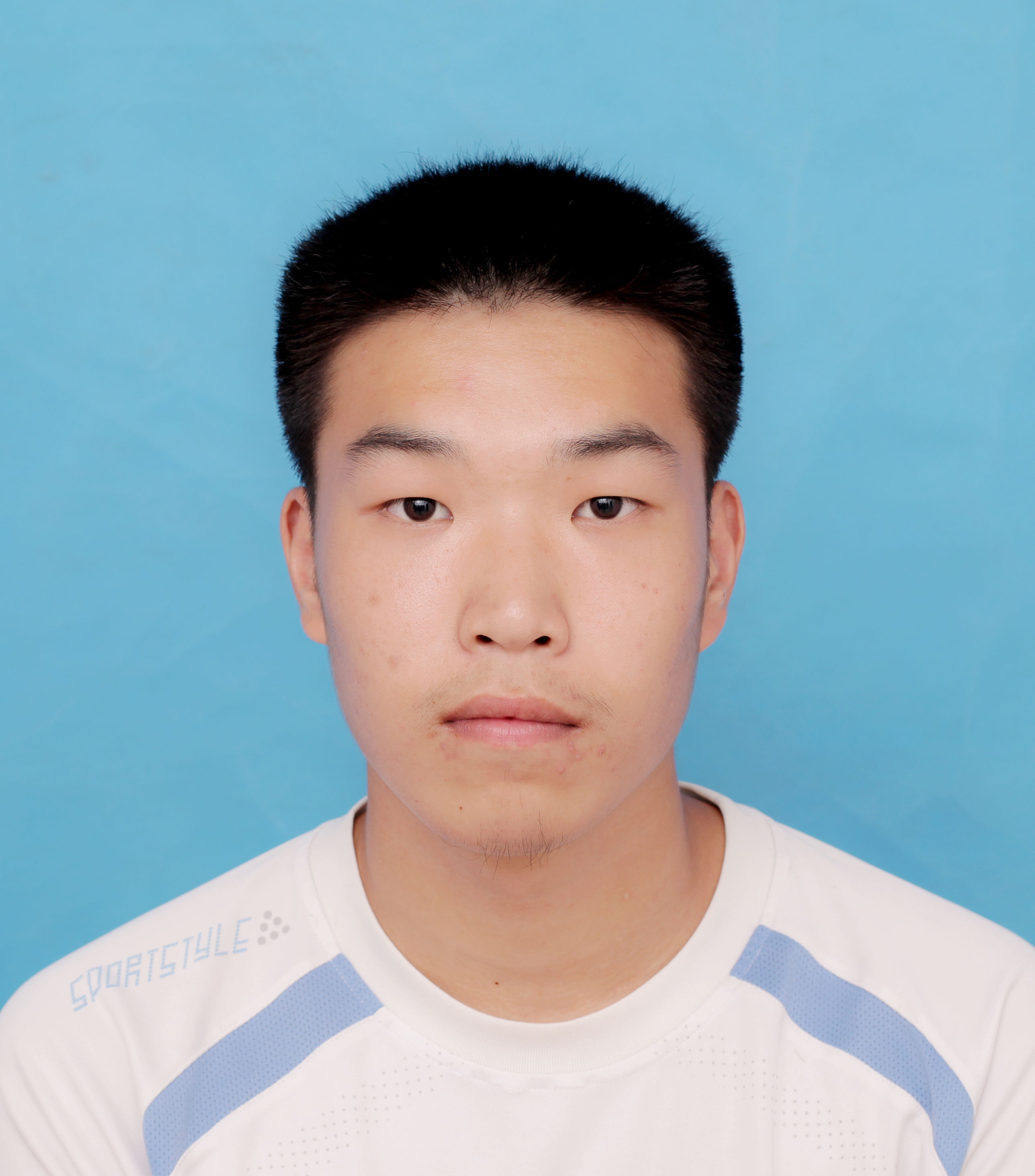}}]{Saichao Liu} received a BS degree and an MS degree in Software Engineering from Henan University, China, in 2019 and 2022, respectively. He is currently pursuing the Ph.D. degree with the College of Computer Science and Technology, Jilin University, Changchun, China. His research interests include wireless communications, UAV networks, antenna arrays and reinforcement learning.
\end{IEEEbiography}

\begin{IEEEbiography}[{\includegraphics[width=1in,height=1.25in,clip,keepaspectratio]{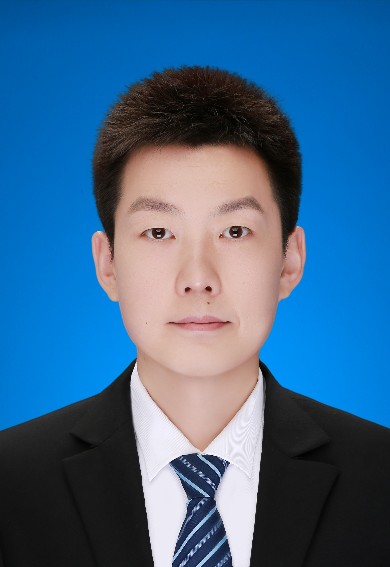}}]{Geng Sun} (S'17-M'19) received the B.S. degree in communication engineering from Dalian Polytechnic University, and the Ph.D. degree in computer science and technology from Jilin University, in 2011 and 2018, respectively. He was a Visiting Researcher with the School of Electrical and Computer Engineering, Georgia Institute of Technology, USA. He is an Associate Professor in College of Computer Science and Technology at Jilin University, and His research interests include wireless networks, UAV communications, collaborative beamforming and optimizations.
\end{IEEEbiography}

\begin{IEEEbiography}[{\includegraphics[width=1in,height=1.25in,clip,keepaspectratio]{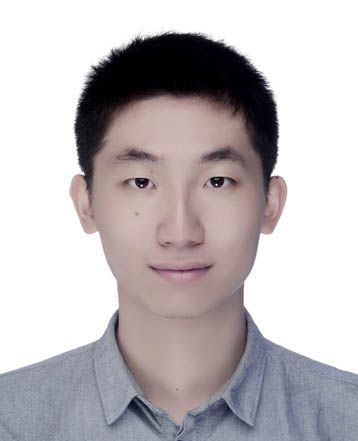}}]{Jiahui Li} (S'21) received a BS degree in Software Engineering, and an MS degree in Computer Science and Technology from Jilin University, Changchun, China, in 2018 and 2021, respectively. He is currently studying Computer Science at Jilin University to get a Ph.D. degree, and also a visiting Ph. D. at Singapore University of Technology and Design (SUTD), Singapore. His current research focuses on UAV networks, antenna arrays, and optimization.
\end{IEEEbiography}

\begin{IEEEbiography}[{\includegraphics[width=1in,height=1.25in,clip,keepaspectratio]{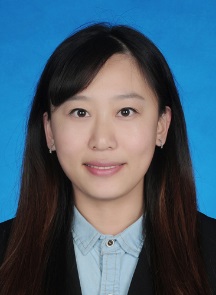}}]{Shuang Liang} received the B.S. degree in Communication Engineering from Dalian Polytechnic University, China in 2011, the M.S. degree in Software Engineering from Jilin University, China in 2017, and the Ph.D. degree in Computer Science from Jilin University, China in 2022. She is a post-doctoral in the School of Information Science and Technology, Northeast Normal University, and her research interests focus on wireless communication and UAV networks.
\end{IEEEbiography}

\begin{IEEEbiography}[{\includegraphics[width=1in,height=1.25in,clip,keepaspectratio]{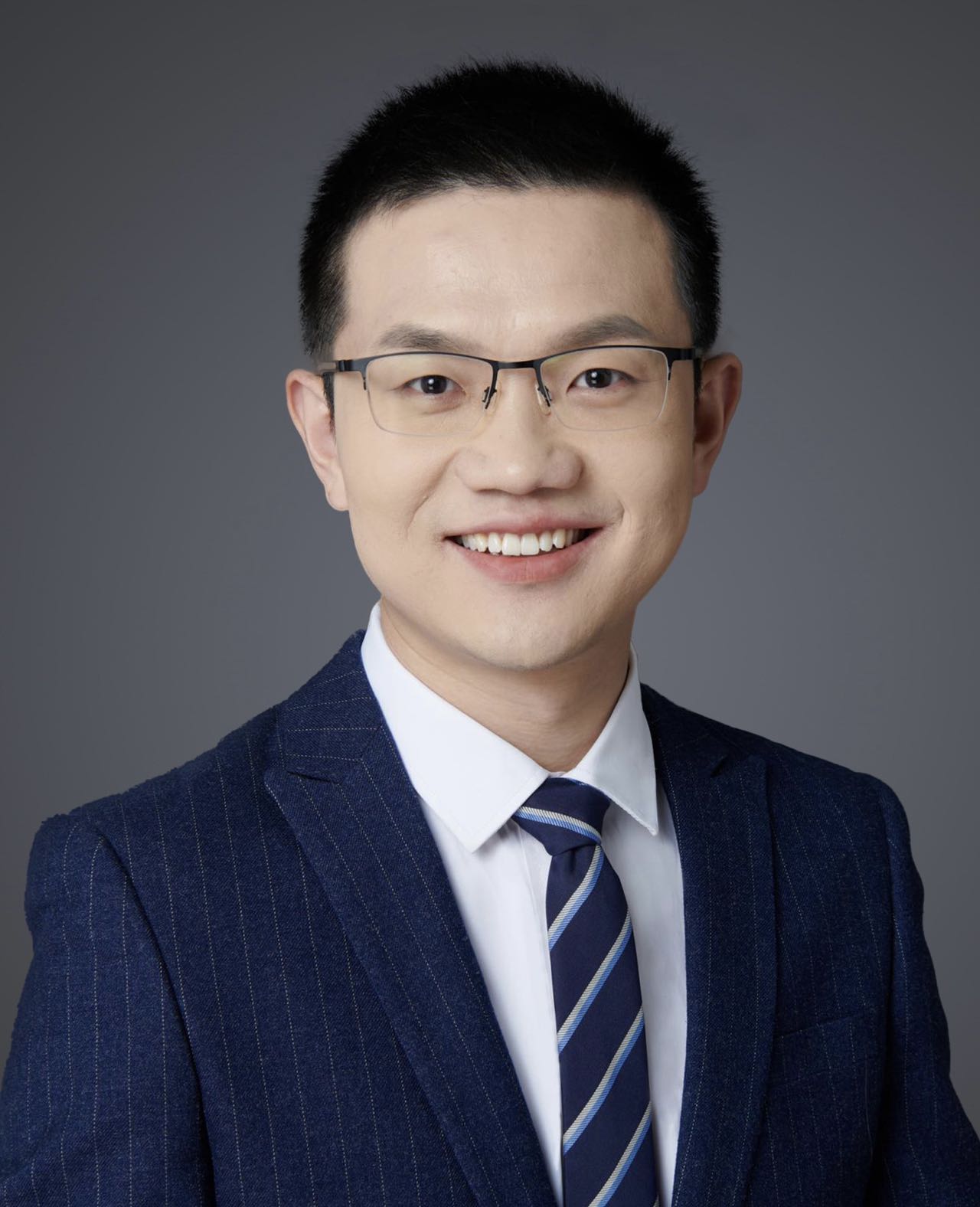}}]{Qingqing Wu} (S’13-M’16-SM’21) received the B.Eng. and the Ph.D. degrees in Electronic Engineering from South China University of Technology and Shanghai Jiao Tong University (SJTU) in 2012 and 2016, respectively. From 2016 to 2020, he was a Research Fellow in the Department of Electrical and Computer Engineering at National University of Singapore. He is currently an Associate Professor with Shanghai Jiao Tong University. His current research interest includes intelligent reflecting surface (IRS), unmanned aerial vehicle (UAV) communications, and MIMO transceiver design. He has coauthored more than 100 IEEE journal papers with 26 ESI highly cited papers and 8 ESI hot papers, which have received more than 18,000 Google citations. He was listed as the Clarivate ESI Highly Cited Researcher in 2022 and 2021, the Most Influential Scholar Award in AI-2000 by Aminer in 2021 and World’s Top 2\% Scientist by Stanford University in 2020 and 2021.
\end{IEEEbiography}

\begin{IEEEbiography}[{\includegraphics[width=1in,height=1.25in,clip,keepaspectratio]{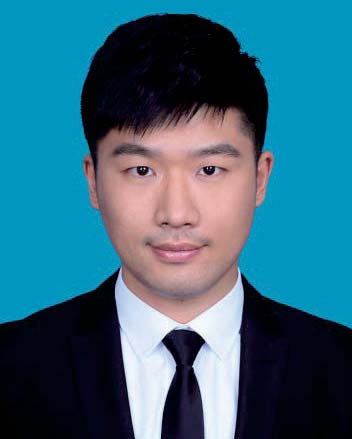}}]{Pengfei Wang} (Member, IEEE) received the B.S., M.S., and Ph.D. degrees in software engineering from Northeastern University (NEU), China,  in 2013, 2015, and 2020, respectively. From 2016 to 2018, he was a Visiting Ph.D. Student with the Department of Electrical Engineering and Computer Science, Northwestern University, IL, USA. He is currently an Associate Professor with the School of Computer Science and Technology, Dalian University of Technology (DUT), China. He has authored more than 30 papers on high-quality journals and conferences, such as IEEE/ACM TRANSACTIONS ON NETWORKING, IEEE INFOCOM, IEEE TRANSACTIONS ON INTELLIGENT TRANSPORTATION SYSTEMS, DAC, IEEE ICNP, IEEE ICDCS, IEEE INTERNET OF THINGS JOURNAL, and JSA. He also holds a series of patents in U.S. and China. His research interests are ubiquitous computing, big data,  and AIoT.
\end{IEEEbiography}


\begin{IEEEbiography}[{\includegraphics[width=1in,height=1.25in,clip,keepaspectratio]{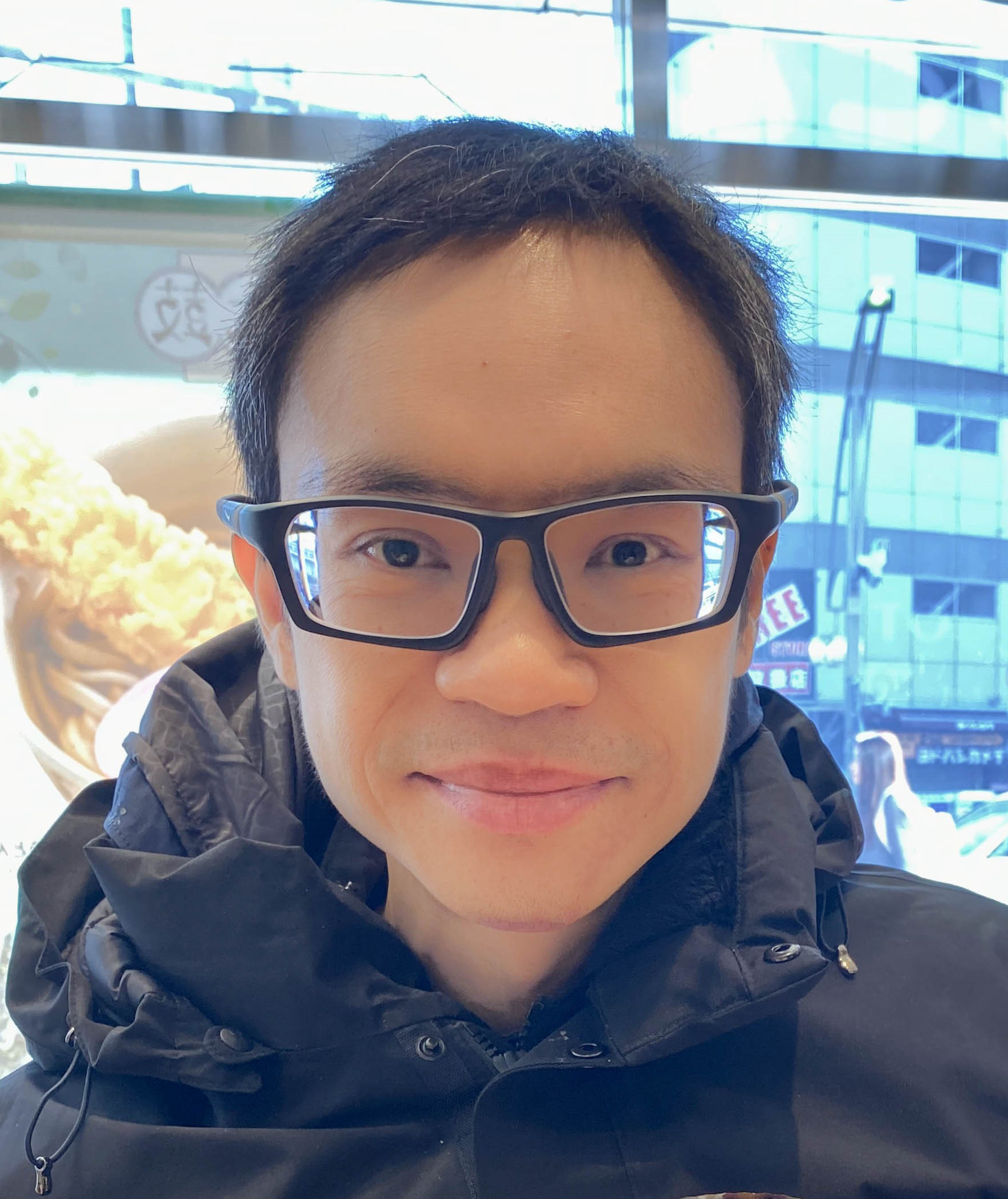}}]{Dusit Niyato} (Fellow, IEEE) received the B.Eng. degree from the King Mongkuts Institute of Technology Ladkrabang (KMITL), Thailand, in 1999, and the Ph.D. degree in electrical and computer engineering from the University of Manitoba, Canada, in 2008. He is currently a Professor with the School of Computer Science and Engineering, Nanyang Technological University, Singapore. His research interests include the Internet of Things (IoT), machine learning, and incentive mechanism design.
\end{IEEEbiography}








\end{document}